\documentclass[twocolumn,superscriptaddress,amsmath,amssymb,notitlepage,nofootinbib]{revtex4-1}

\usepackage{amsthm}
\usepackage{amsmath}
\usepackage{amsfonts}
\usepackage{amssymb}
\usepackage{graphicx}
\usepackage{epstopdf}
\usepackage{mathrsfs}
\usepackage{color}
\usepackage{chngcntr}
\usepackage{tabularx}
\usepackage{booktabs}
\usepackage{lineno}
\usepackage{mathrsfs}
\usepackage[mathscr]{euscript}
\usepackage{bm}

\counterwithout{figure}{section}
\numberwithin{equation}{section}

\newcommand{\la}{\vec{\lambda}}
\newcommand{\crema}{CReM$_A$}
\newcommand{\cremb}{CReM$_B$}

\begin{document}

\title{A faster horse on a safer trail:\\ generalized inference for the efficient reconstruction of weighted networks}

\author{Federica Parisi}
\affiliation{IMT School for Advanced Studies Lucca}
\author{Tiziano Squartini}
\email{tiziano.squartini@imtlucca.it}
\affiliation{IMT School for Advanced Studies Lucca}
\author{Diego Garlaschelli}
\affiliation{IMT School for Advanced Studies Lucca}
\affiliation{Lorentz Institute for Theoretical Physics, Leiden University}
\date{\today}

\begin{abstract}
Due to the interconnectedness of financial entities, estimating certain key properties of a complex financial system, including the implied level of systemic risk, requires detailed information about the structure of the underlying network of dependencies. However, since data about financial linkages are typically subject to confidentiality, network reconstruction techniques become necessary to infer both the presence of connections and their intensity. Recently, several ``horse races'' have been conducted to compare the performance of the available financial network reconstruction methods. These comparisons were based on arbitrarily chosen metrics of similarity between the real network and its reconstructed versions. Here we establish a generalised maximum-likelihood approach to rigorously define and compare weighted reconstruction methods. Our generalization uses the maximization of a certain conditional entropy to solve the problem represented by the fact that the density-dependent constraints required to reliably reconstruct the network are typically unobserved and, therefore, cannot enter directly, as sufficient statistics, in the likelihood function. The resulting approach admits as input any reconstruction method for the purely binary topology and, conditionally on the latter, exploits the available partial information to infer link weights. We find that the most reliable method is obtained by ``dressing'' the best-performing binary method with an exponential distribution of link weights having a properly density-corrected and link-specific mean value and propose two safe (i.e. unbiased in the sense of maximum conditional entropy) variants of it. While the one named \crema\:is perfectly general (as a particular case, it can place optimal weights on a network if the bare topology is known), the one named \cremb\:is recommended both in case of full uncertainty about the network topology and if the existence of some links is certain. In these cases, the \cremb\:is faster and reproduces empirical networks with highest generalised likelihood among the considered competing models.
\end{abstract}


\maketitle

\section{Introduction}

Network reconstruction is an active field of research within the broader field of complex networks. In general, network reconstruction consists in facing the double challenge of inferring both the bare topology (i.e. the existence or absence of links) and the magnitude (i.e. the weight) of the existing links of a network for which only aggregate or partial structural information is known.
These two pieces of the puzzle (i.e. the `topology' and the `weights') represent equally important targets of the reconstruction problem, although reaching those targets may require very different strategies.
In general, the available pieces of information represent the constraints guiding the entire reconstruction procedure. Depending on the nature of the available constraints, different reconstruction scenarios materialize.
The scenario considered in this paper is the one that is recurrently encountered in the study of financial and economic networks~\cite{squartini2018reconstruction,cimini2019statistical}.

Indeed, financial networks are a class of networks for which the reconstruction challenge is particularly important.
The estimation of systemic risk, the simulation of financial contagion and the `stress testing' of a financial network in principle require the complete knowledge of the underlying network structure.
If the description of this structure is naively simplified or reduced, then the outcome of those stress tests becomes unreliable when taken as a proxy of what would happen on the real network in the same situation.
This may imply a severe underestimation of the level of systemic risk, as research conducted in the aftermath of the 2007-2008 crisis showed.
In the typical situation for financial networks,  the total number $N$ of nodes (e.g. the number of banks in a given network of interbank lending) is known, but the number, intensity and position of the links among those nodes is uknown because of confidentiality issues. 
Generally, one has access only to node-specific information that is publicly available. 
For instance, from publicly reported balance sheets one knows the so-called `total assets' (total value of what a bank owns, including what it lent out to other banks in the network) and `total liabilities' (total value of what a bank owes to the external world, including what it borrowed from other banks in the network) of each bank in an interbank system. 
Similar considerations apply to inter-firm networks, where links represent typically unobservable individual transactions while the total purchases and total sales of each of the firms in the system considered are more easily accessible.
One more example, which is relevant not strictly for the reconstruction problem but rather from a modelling point of view, is that of the international trade network, where one would like to obtain a good model of international trade flows from country-specific aggregate quantities such as the total imports and total exports of each country.

In all the examples mentioned above, the pieces of node-specific information typically represent a good proxy of the \emph{margins}, i.e. the sums along columns and rows, of the weighted adjacency matrix $\mathbf{W}^*$ of the (in general directed) network, whose entry $w^*_{ij}$ quantifies the magnitude of the link existing from node $i$ to node $j$ (including $w^*_{ij}=0$ if no link is there). 
In the language of network science, these two margins are called the \emph{out-strength} $s^{out^*}_i\equiv \sum_{j\ne i}w_{ij}^*$ and the \emph{in-strength} $s^{in^*}_i\equiv\sum_{j\ne i}w_{ji}^*$ of node $i$, where the asterisk indicates the `true' value, i.e. the value measured on the true network $\mathbf{W}^*$, of those quantities.
In general, one assumes that the full matrix $\mathbf{W}^*$ itself is unobservable, while $s^{out^*}_i$ and $s^{in^*}_i$ are (directly or indirectly) measurable for each node ($i=1\dots N$). The $N$-dimensional vectors $\vec{s}^{out^*}$ and $\vec{s}^{in^*}$ constructed from all node strengths are called the \emph{out-strength sequence} and \emph{in-strength sequence} respectively.
It is worth stressing here that the in- and out-strength sequences represent a form of \emph{weighted} constraints that can be imposed in the reconstruction procedure, because they depend explicitly on the magnitude of the links in the network. 
As such, they do not contain direct information about the \emph{binary topology} of the network, such as the overall density of links, the number of links (i.e the \emph{degree}) of each node, etc.
This makes the simultaneous inference of both the link weights \emph{and} the bare topology of the network particularly challenging in this setting.

Irrespective of how the strength sequences are used in the reconstruction method, it is clear that there are multiple (in fact, hugely many) possible networks that are consistent with such margins. 
The essence of each method lies in how this set of compatible networks is further restricted to produce the output networks.
At one extreme, there are greedy methods based on certain heuristics or ansatz that in the end produce a single possible instance of the network. 
The problem with these `deterministic' methods is that, by producing a single outcome, they give zero probability to any other network, including (apart from sheer luck) the true unobserved network $\mathbf{W}^*$. This implies that the likelihood of producing the real network given the model is always zero. The success of such deterministic methods, as well as their comparison with competing methods, has therefore to be assessed via some arbitrarily chosen metric of network similarity.
At the other extreme, there are maximally agnostic methods designed to impose absolutely no other ansatz or heuristic besides the knowledge of the strengths sequences, so that all the compatible networks are accepted with equal probability.
This is the class of (unconditional) \emph{maximum-entropy methods}, that look for the probability distribution (in the space of weighted networks) maximizing the entropy, subject to the imposed constraints. 
Research has shown that typical networks sampled from maximum-entropy ensembles of networks with given strength sequence are fully or almost fully connected \cite{mastrandrea2014enhanced}.
In light of the sparsity of most real-world networks, this is a major limitation.

All the state-of-the-art reconstruction methods are found somewhere in between the two extreme cases described above. 
Among the methods proposed so far, some assume that the constraints concerning the binary and the weighted network structure jointly determine the reconstruction output. 
An example providing an excellent reconstruction of several real-world weighted networks is the Enhanced Configuration Model (ECM) \cite{mastrandrea2014enhanced}, defined by simultaneously constraining the nodes degrees and strengths.
However the inaccessibility of empirical degrees makes this method inapplicable in the setting considered here.
This has led to the introduction of two-step algorithms \cite{cimini2015systemic,cimini2015estimating} that perform a preliminary estimation of node degrees to overcome the lack of binary information. 
Other methods consider the weights estimation step as completely unrelated to the binary one \cite{andrecut2016systemic,halaj2013assessing}. Examples include methods that adjust the link weights iteratively on top of some previously determined topological structure (e.g. via the recipe firstly proposed in \cite{bacharach1965estimating}), in such a way to satisfy the strengths constraints \emph{a posteriori}. This kind of procedure, however, assigns weights deterministically, thus being unable to provide confidence bounds accompanying the weight estimates \cite{gandy2016bayesian} and basically giving zero probability to any real-world network.

In this paper, after reviewing the state-of-the-art methods and discussing their performance, we develop a theoretical framework that provides an analytical, unbiased\footnote{Throughout the paper, we use the term `unbiased' as intended in the application of the maximum-entropy principle, i.e. when we refer to outcomes that maximize the (conditional) entropy, so that the resulting probability distribution does not make arbitrary preferences (corresponding to hidden or unjustified assumptions) among configurations that share the same values of certain target quantity. Constrained maximum-entropy distributions produce maximally random outcomes given what is supported by the data used as constraints, thereby ensuring unbiasedness. To avoid confusion with the meaning of the term `bias' in statistics, we do not use the term in the sense of `biased parameter estimation'.}
 (i.e. based on the maximization of a certain conditional entropy) procedure to estimate the weighted structure of a network.
The maximization of the conditional entropy generalizes the Exponential Random Graph (ERG) formalism to situations where the aggregate topological properties that effectively summarize the network topology are not directly observable and cannot therefore enter as sufficient statistics into the model (and in the ordinary likelihood function). 
Information about the topological structure (either available \emph{ab initio} or obtained by using any of the existing algorithms for the purely binary reconstruction) is treated as \emph{prior} information.
Together with the available weighted constraints, this prior information represents the input of our generalized reconstruction procedure. 
The probability distribution describing link weights is, then, determined by maximizing a suitably defined conditional entropy.
This construction allows us to achieve an optimal compromise between the deterministic and fully agnostic extremes mentioned above: while we allow the method to incorporate a certain ansatz (both for the purely binary structure and for the weights) that effectively restricts the set of compatible networks, we still maximize a certain entropy in order to preserve the necessary indifference among configurations that have the same `good' properties, induced by the ansatz itself. Finally, the parameters of the conditionally maximum-entropy distribution are found by maximizing a generalized likelihood function that depends on the probability distribution over binary graphs implied by the binary reconstruction method. This last step makes the weight distribution dependent, as it should, on the purely binary expected network properties.

As it turns out, when link weights are treated as continuous random variables, their distribution - conditional on the existence of the links themselves - is exponential, a result that can be used to further enhance the performance of the best-performing methods available to date, providing them with a recipe to determine confidence intervals for the weight estimates. While it is a well know result that the exponential distribution follows from the maximization of the entropy subject to a constraint on the mean value, what is nontrivial here is the determination of how the mean value itself should depend on a combination of certain empirically observed regularities and, crucially, on the prior probability distribution of the bare topological projection of the network implied by the binary reconstruction method chosen as input. As a byproduct, our generalized reconstruction framework leads to a computationally simpler variant of our method, based on the solution of a single nonlinear equation in place of several coupled, non-linear equations as in some of the previous methods\footnote{Where appropriate, the interested reader will be redirected towards freely available codes to run all variants of our framework.}.

The rest of the paper is organized as follows. In section \ref{sec:methods} we review the state of the art of network reconstruction methods and discuss their performance. We then describe our generalized `conditional reconstruction method' in detail, providing two different specifications of it. In section \ref{sec:results} we test the performance of the method on real-world networks. In section \ref{sec:4} we discuss the results.

\section{Methods\label{sec:methods}}

In what follows, we indicate a weighted adjacency matrix as $\mathbf{W}$ and its generic entry as $w_{ij}$. Analogously, we indicate the corresponding adjacency matrix as $\mathbf{A}$ and its entry as $a_{ij}=\Theta(w_{ij})$, with $\Theta(x)$ representing the Heaviside step function, defined as $\Theta(x)=1$ if $x>0$ and $\Theta(x)=0$ if $x\le 0$.

\subsection*{Network reconstruction methods:\\an overview of the state-of-the-art}

\paragraph*{The MaxEnt method: deterministic link weights on a complete graph.}
A traditional approach to network reconstruction is the so-called MaxEnt method \cite{wells2004financial,upper2011simulation,mistrulli2011assessing}, defined by the maximization of the `entropy' $S(\mathbf{W})=-\sum_{i,j}w_{ij}\ln w_{ij}$  under the constraints represented by the network weighted marginals, i.e. $s_i^{out^*}=\sum_{j\neq i}w^*_{ij},\:\forall\:i$ and $s_i^{in^*}=\sum_{j\neq i}w^*_{ji},\:\forall\:i$. 
The resulting `maximum-entropy' expression for $w_{ij}$ is easily found to be
\begin{equation}
\hat{w}^{\text{ME}}_{ij}=\frac{s_i^{out^*}s_j^{in^*}}{W^*},\quad\forall\:i,j
\label{eq:cw_gm}
\end{equation}
with $W^*=\sum_is_i^{out^*}=\sum_is_i^{in^*}$.
The major drawback of the above model is the prediction of a fully connected network with all positive link weights given by eq. \eqref{eq:cw_gm}. 
Yet, the above expression often provides an accurate estimation of the subset of realized (i.e. positive) real-world link weights. This fact will turn out useful later in our analysis.
At a fundamental level, the ultimate issue with this method is that, although the quantity $S(\mathbf{W})$ is referred to as `entropy', actually the link weight $w_{ij}$ admits no natural interpretation as a probability distribution over the entries of the matrix, contrary to what the definition of entropy would instead require.
In particular, $S(\mathbf{W})$ is a function of a \emph{single} matrix, rather than a function $S(\mathcal{W})$ of a \emph{probability distribution} over an ensemble of realizations of the matrix, treated as a random variable $\mathcal{W}$ that can take $\mathbf{W}$ as one of its possible values with a certain probability (the approach that we introduce later will be based precisely on a proper entropy $S(\mathcal{W})$ of this type, and particularly on a certain conditional version of it).
This consideration immediately questions the interpretation of eq. \eqref{eq:cw_gm} as a truly `maximum-entropy' result.
In fact, by producing a single matrix as output, the method is actually a deterministic (i.e. a zero-entropy) one, rather than a probabilistic one as proper maximum-entropy methods necessarily are.\\

\paragraph*{Iterative Proportional Fitting: deterministic link weights on any graph.} The search for nontrivial (i.e. sparser) topological configurations still guaranteeing that the weighted marginals are satisfied has led to a plethora of reconstruction methods. These models are described below; here we mention an aspect common to many of them. Irrespective of the method used for the reconstruction of the binary topology, a popular way to assign link weights on a non-complete (not fully connected) graph, while still matching the constraints given by the in- and out-strength sequences, is the Iterative Proportional Fitting (IPF) algorithm \cite{bacharach1965estimating}. The IPF recipe assumes that the network topology is given and iteratively ``adjusts'' link weights until the constraints are satisfied \cite{squartini2018reconstruction,bacharach1965estimating}. In the special case when the network is fully connected, the IPF algorithm reduces to MaxEnt. Since the IPF algorithm always yields a (unique) matrix satisfying the weighted marginals irrespective of the topological details\footnote{The only request about $\mathbf{A}$ is its irreducibility \cite{squartini2018reconstruction,bacharach1965estimating}.} of the given underlying binary structure $\mathbf{A}$, many researchers have focused on methods for improving the reconstruction of the bare network topology, while considering the `link weight' problem virtually solved and, more importantly, decoupled from the `topology' problem. As we argue later on, this consideration is incorrect. Moreover, the IPF recipe suffers from two serious drawbacks, both imputable to the deterministic rule used to assign weights to a given binary configuration. First, it cannot provide confidence bounds accompanying the weight estimates. Second, the probability of reproducing any real-world weighted network is virtually zero, even if the bare topology were known exactly.\\

\paragraph*{Many horses and many races.} Here we succintly describe the state-of-the-art reconstruction methods (the ``horses'') that have been recently considered in various ``horse races'' comparing the performance of different methods over a number of real-world networks.
These methods have been recently reviewed in \cite{squartini2018reconstruction} and are here compactly collected in table \ref{tab_meth}. In order to unambiguously assess the performance of a given method, we consider the probability (density) $Q(\mathbf{W})$ of generating a given weighted graph $\mathbf{W}$ according to the method, and use the corresponding log-likelihood
\begin{eqnarray}\label{score}
\ln Q(\mathbf{W}^*)&=&\ln[P(\mathbf{A}^*)Q(\mathbf{W}^*|\mathbf{A}^*)]\nonumber\\
&=&\ln P(\mathbf{A}^*)+\ln Q(\mathbf{W}^*|\mathbf{A}^*)
\end{eqnarray}
as a score function quantifying how likely the structure of the specific real-world network $\mathbf{W}^*$ is reproduced by a given algorithm. Notice that we have written $Q(\mathbf{W})=P(\mathbf{A})Q(\mathbf{W}|\mathbf{A})$ where $P(\mathbf{A})$ is the probability of generating the bare topology $\mathbf{A}$ of $\mathbf{W}$ and $Q(\mathbf{W}|\mathbf{A})$ is the conditional probability of generating the weights of the network, given its topology.
Therefore $P(\mathbf{A}^*)$ is a sort of purely binary likelihood.

Upon looking at table \ref{tab_meth}, several classes of algorithms can be distinguished. A first group gathers the algorithms whose estimation steps are both deterministic (notice that the purely deterministic version of the minimum-density algorithm is considered here, i.e. the one used in the ``horse race'' \cite{anand2018missing}). Since the deterministic nature of an algorithm implies that the probability $f_{ij}$ that nodes $i$ and $j$ are connected is $f_{ij}\in\{0,1\}$, the only way to reproduce the actual topological structure $\mathbf{A}^*$ is implementing the rule $f_{ij}=1\iff a_{ij}^*=1$ and $f_{ij}=0\iff a_{ij}^*=0$. However, this prescription is viable only if the actual configuration is given, otherwise the probability of reproducing its structure is $P(\mathbf{A}^*)=0$, further impliying that $\ln Q(\mathbf{W}^*)=-\infty$.

A second group gathers algorithms where the topological estimation step is indeed probabilistic while the recipe for assigning link weights is deterministic: while the vast majority of such methods rests upon the IPF algorithm, the \emph{density-corrected Gravity Model (dcGM)} \cite{cimini2015systemic} method rests upon the MaxEnt prescription. The method proposed in \cite{gandy2016bayesian}, instead, employs the \emph{maximum-flow (MF)} algorithm to adjust weights. Even if these algorithms indeed allow for the observed topological structure to be replicable (i.e. $P(\mathbf{A}^*)>0$), they still assign weights in a deterministic fashion: this implies that the actual configuration $\mathbf{W}^*$ can be reproduced if and only if $\hat{w}_{ij}=w_{ij}^*$, i.e. only in case the actual configuration is given, otherwise $Q(\mathbf{W}^*|\mathbf{A}^*)=0$, again impliying $\ln Q(\mathbf{W}^*)=-\infty$.

A third group gathers algorithms whose steps are both probabilistic. However, weights are assumed to be natural numbers: hence, configurations with \emph{real} weights - i.e. typical real-world networks - cannot, by definition, be reproduced by such recipes.

The last two methods may, in principle, lead to recover the structure of a network with real-valued weights. However, the method by Ha\l{}aj \& Kok induces a completely random topological structure, leading to a probability of reproducing any observed $\mathbf{A}^*$ that reads $P(\mathbf{A}^*)=2^{-N(N-1)}$, rapidly vanishing as $N$ grows.
On the other hand, the method proposed by Moussa aims at reproducing a specific feature of several real-world networks, i.e. a power-law degree distribution: as a consequence, it is optimized to reconstruct such a peculiar topological feature and does not allow for generic degree distributions. Moreover, the method does not come with a recipe for assigning the generated degrees to the different nodes.\\

\begin{table}[t!]\footnotesize
\begin{tabular}{cccc}
\hline
\hline
{\bf Method} & {\bf Topology} & {\bf Weights} & {\bf Log-likelihood} \\
\hline
\hline
MaxEnt (ME) \cite{wells2004financial,upper2011simulation} & D & D & $-\infty$ \\
\addlinespace
Minimum-Density (MD) \cite{anand2015filling} & D & D & $-\infty$ \\
\hline
Copula approach \cite{baral2012estimating} & P & D (IPF) & $-\infty$ \\
\addlinespace
Drehmann \& Tarashev \cite{drehmann2013measuring} & P & D (IPF) & $-\infty$ \\
\addlinespace
Montagna \& Lux \cite{montagna2017contagion} & P & D (IPF) & $-\infty$ \\
\addlinespace
Mastromatteo et al. \cite{mastromatteo2012reconstruction} & P & D (IPF) & $-\infty$ \\
\addlinespace
Gandy \& Veraart \cite{gandy2016bayesian} & P & D (MF) & $-\infty$ \\
\addlinespace
dcGM \cite{cimini2015systemic} & P & D (ME) & $-\infty$ \\
\hline
MECAPM \cite{digiangi2016assessing} & P & P ($w_{ij}\in\mathbb{N}$) & $-\infty$ \\
\addlinespace
fitness-induced DECM \cite{cimini2015estimating} & P & P ($w_{ij}\in\mathbb{N}$) & $-\infty$ \\
\hline
Ha\l{}aj \& Kok \cite{halaj2013assessing} & P & P & $\in\mathbb{R}$ \\
\addlinespace
Moussa \cite{moussa2011contagion} & P & P & $\in\mathbb{R}$ \\
\hline
\hline
\end{tabular}
\caption{Overview of the reconstruction methods reviewed in \cite{squartini2018reconstruction}. The letter ``P'' indicates that the considered estimation step is probabilistic in nature while the letter ``D'' indicates that it is deterministic. The log-likelihood is defined as in eq. (\ref{score}), i.e. $\ln Q(\mathbf{W}^*)$.}\label{tab_meth}
\end{table}

\paragraph*{A good horse on the binary trail: the density-corrected gravity model.} The above considerations imply that, as far as the simultaneous reconstruction of both the topology and the weights is concerned, none of the current methods is satisfactory. The remainder of the paper aims at introducing a viable and efficient method. The method will be designed in such a way that any purely binary reconstruction method, i.e. any $P(\mathbf{A})$, can be taken as input, while aiming at placing link weights optimally. This will allow us to freely choose the binary method at the end. It is therefore worthwhile to describe in some detail here the specific binary method that we will eventually select for our analyses when putting the full method at work.
Our choice is guided by the results of four independent tests (``horse races'') \cite{anand2018missing,mazzarisi2017methods,ramadiah2017reconstructing,lebacher2019search}, that have found that, as far as the imputation of the overall binary topology of the network is concerned, the dcGM \cite{cimini2015systemic} systematically outperforms competing methods. Quoting from the source references:

\begin{itemize}
\item ``\emph{In presenting our results we face the challenge that some algorithms produce an ensemble of networks while others produce a single matrix. This makes a straightforward comparison difficult. Fortunately, the Cimi method is the clear winner between the ensemble methods}'' \cite{anand2018missing} (note: `Cimi' is the name given in \cite{anand2018missing} to the dcGM); 
\item ``\emph{According to our analysis, reconstructing via fitness model outperforms the other methods when the same input information is used}'' \cite{mazzarisi2017methods} (note: `fitness model' is the name given in \cite{mazzarisi2017methods} to the dcGM);
\item ``\emph{Second, concerning the individual performance of each null model, we find that CM1, followed by CM2 and MaxEntropy, has the closest behavior to the actual network overall. Since CM2 requires much less information than CM1, we find that this makes CM2 more appealing for practical purposes}'' \cite{ramadiah2017reconstructing} (note: `CM2' is the name given in \cite{ramadiah2017reconstructing} to the dcGM);
\item ``\emph{As an `off the shelf' model in situations without exogenous information available, the density-corrected gravity model (DC-GRAVITY) can be recommended because it is found to work well on the big sparse network as well as on the small dense network with respect to the edge probabilities and the edge values [\dots] Similarly, Gandy and Veraart (2019) report that this model is performing very well in binary and valued reconstruction. Further, the model can be extended towards the inclusion of exogenous information in a simple way}'' \cite{lebacher2019search} (note: `DC-GRAVITY' is the name given in \cite{ramadiah2017reconstructing} to the dcGM).
\end{itemize}

The above results prompt us to select the binary part of the dcGM as the best candidate to be given as input to the full method to be developed.
The dcGM is defined by the simple Bernoulli prescription
\begin{eqnarray}
a_{ij}=\begin{cases}
1 & \mbox{with\:\:\:\:} p_{ij}^{\text{dcGM}}=\frac{zs_i^{out}s_j^{in}}{1+zs_i^{out}s_j^{in}}\\
0 & \mbox{with\:\:\:\:} 1-p_{ij}^{\text{dcGM}}
\end{cases}
\end{eqnarray}
(for $i\neq j$), where the only free parameter $z$ is tuned to reproduce the actual link density of the network \cite{cimini2015systemic}. It is worth mentioning here that the dcGM takes the functional form of the connection probability from the binary configuration model (BCM), i.e. the maximum-entropy model of binary graphs with given in- and out-degrees for all nodes (see Appendix). In the BCM, the parameters are Lagrange multipliers that control the in- and out-degrees. In the dcGM, these parameters (that are unidentifiable, given the inaccessibility of the degrees) are replaced by the observed values of the in- and out-strengths respectively, up to the global proportionality constant $z$. This is the so-called `fitness ansatz', motivated by an empirical regularity showing the systematic approximate proportionality between empirical strenghts and Lagrange multipliers coupled to the degrees \cite{cimini2015systemic.
More details are provided in the Appendix.}

\subsection*{A framework for conditional reconstruction}\label{sec:1}

In the following, we aim at introducing a method overcoming the drawbacks affecting current weight imputation recipes. Ideally, our recipe should satisfy the following requirements:

\begin{itemize}
\item allowing for any probability distribution (over purely binary graphs) to be acceptable as input for the preliminary topology reconstruction step (this requirement allows us to take any binary reconstruction method as input - and clearly, to select a good one for practical purposes);
\item allowing for the generation of continuous weights (this requirement implies that the real unobserved network will be generated with positive likelihood);
\item satisfying the constraints that are usually imposed by the availability of limited information (i.e. the out- and in-strength sequences $\{s_i^{out}\}_{i=1}^N$ and $\{s_i^{in}\}_{i=1}^N$).
\end{itemize}

As anticipated in the Introduction, these three postulates will be addressed by proposing a probabilistic reconstruction method \emph{conditional on some prior binary information} and constrained to reproduce the aforementioned, weighted observables. In order to do so, we build upon the formalism proposed by the authors of \cite{gabrielli2018grand} who define a fully probabilistic procedure to \emph{separately} constrain binary and weighted network properties. In short, they introduced the continuous version of the ECM \cite{mastrandrea2014enhanced} and replaced the resulting probability of the binary projection of the network with the one coming from the Undirected Binary Configuration Model (UBCM) \cite{squartini2015unbiased}. In such a way, the estimation of the probability coefficients controlling for the presence of links is disentangled from the estimation step concerning link weights. 
Unfortunately, the framework proposed in \cite{gabrielli2018grand} cannot be directly used for network reconstruction as the information about the degrees of nodes is practically never accessible.

\section{Results\label{sec:results}}

Before entering into the details of our method, let us briefly describe the formalism we adopt. In what follows, we assume that $\mathbf{A}\in\mathbb{A}$ is a realization of the random variable $\mathbf{\mathcal{A}}$; analogously, the weighted adjacency matrix $\mathbf{W}\in\mathbb{W}$ instantiates the random variable $\mathbf{\mathcal{W}}$. The probability mass function of the event $\mathbf{\mathcal{A}}=\mathbf{A}$ is denoted with $P(\mathbf{A})$, while $Q(\mathbf{W}|\mathbf{A})$ is a conditional probability density function, for the variable $\mathcal{W}$ taking the value $\mathbf{W}$, given the event $\mathbf{\mathcal{A}}=\mathbf{A}$. Notice that we are considering continuous (non-negative and real-valued) weights and that $Q(\mathbf{W}|\mathbf{A})$ is non-zero only over the continuous set $\mathbb{W}_\mathbf{A}=\{\mathbf{W}: \Theta(\mathbf{W})=\mathbf{A}\}$ of weighted matrices with binary projection equal to $\mathbf{A}$.\\

\paragraph*{Input.} Our Conditional Reconstruction Method (CReM) takes as input $P(\mathbf{A})$, i.e. the distribution over the space of binary configurations: this is treated as prior information and can be computed by using any available method.
Clearly, given the superior performance of the dcGM as summarized above, we will select that particular model in our own analysis, but nonetheless we want to keep the method as general as possible by allowing for any input $P(\mathbf{A})$. Moreover, the CReM requires as input a set of weighted constraints $\vec{C}(\mathbf{W})$ representing the available information about the system at hand. The observed numerical value of these constraints will be denoted by $\vec{C}^*$. The true, unobserved matrix will be denoted with $\mathbf{W}^*$ and the associated binary projection with $\mathbf{A}^*$. Clearly $\vec{C}(\mathbf{W}^*)=\vec{C}^*$.\\

\paragraph*{Output.} The goal of the CReM is to derive the distribution over the ensemble of weighted configurations conditional on the prior information concerning the binary ensemble. To this aim, we look for a functional form of $Q(\mathbf{W}|\mathbf{A})$ such that $Q(\mathbf{W}|\mathbf{A})=0$ for $\mathbf{W}\notin\mathbb{W}_\mathbf{A}$ and otherwise maximizing the \emph{conditional entropy} \cite{cover2006elements}

\begin{equation}\label{eq:condent}
S(\mathbf{\mathcal{W}}|\mathbf{\mathcal{A}})=-\sum_{\mathbf{A}\in\mathbb{A}}P(\mathbf{A})\int_{\mathbb{W}_\mathbf{A}}Q(\mathbf{W}|\mathbf{A})\log Q(\mathbf{W}|\mathbf{A})d\mathbf{W}
\end{equation}
under the set of constraints

\begin{eqnarray}
1&=&\int_{\mathbb{W}_\mathbf{A}}Q(\mathbf{W}|\mathbf{A})d\mathbf{W},\:\forall\:\mathbf{A}\in\mathbb{A}\label{cone}\\
\langle C_\alpha\rangle&=&\sum_{\mathbf{A}\in\mathbb{A}} P(\mathbf{A})\int_{\mathbb{W}_\mathbf{A}}Q(\mathbf{W}|\mathbf{A})C_{\alpha}(\mathbf{W})d\mathbf{W}=C_{\alpha}^*,\:\forall\:\alpha\label{ctwo}\nonumber\\
\end{eqnarray}
(notice that $\langle\cdot\rangle$ denotes an average with respect to $Q(\mathbf{W})$). Equation (\ref{cone}) defines the normalization of the conditional probability and ensures that the unconditional probability 

\begin{equation}
Q(\mathbf{W})=\sum_{\mathbf{A}\in\mathbb{A}}P(\mathbf{A})Q(\mathbf{W}|\mathbf{A})
\end{equation}
is also normalized as 

\begin{equation}
\int_\mathbb{W}Q(\mathbf{W})d\mathbf{W}\equiv\sum_{\mathbf{A}\in\mathbb{A}}\int_{\mathbb{W}_\mathbf{A}}Q(\mathbf{W})d\mathbf{W}=\sum_{\mathbf{A}\in\mathbb{A}}P(\mathbf{A})=1;\nonumber\\
\end{equation}
equation (\ref{ctwo}), instead, sets the target values $\vec{C}^*$ of our constraints. The problem Lagrangean can be, thus, written as the following generalization of the Lagrangean valid in the unconditional case (see also the Appendix):

\begin{eqnarray}\label{eq:lagr}
\mathscr{L}&=&S(\mathbf{\mathcal{W}}|\mathbf{\mathcal{A}})\nonumber\\
&&+\sum_{\mathbf{A}\in\mathbb{A}}\mu(\mathbf{A}) \left(1-\int_{\mathbb{W}_\mathbf{A}}Q(\mathbf{W}|\mathbf{A})d\mathbf{W}\right)\nonumber\\
&&+\sum_{\alpha}\lambda_{\alpha}\left(C_{\alpha}^*-\sum_{\mathbf{A}\in\mathbb{A}}P(\mathbf{A})\int_{\mathbb{W}_\mathbf{A}}Q(\mathbf{W}|\mathbf{A})C_{\alpha}(\mathbf{W})d\mathbf{W}\right).\nonumber\\
\end{eqnarray}
Differentiating with respect to $Q(\mathbf{W}|\mathbf{A})$ and equating the result to zero leads to

\begin{equation}\label{eq:q_exprg}
Q_{\la}(\mathbf{W}|\mathbf{A})=\left\{\begin{array}{cc}
\frac{e^{-H_{\la}(\mathbf{W})}}{Z_{\mathbf{A},\la}}&\mathbf{W}\in\mathbb{W}_\mathbf{A}\\
0&\mathbf{W}\notin\mathbb{W}_\mathbf{A}
\end{array}\right.
\end{equation}
where $H_{\la}(\mathbf{W})=\sum_{\alpha}\lambda_{\alpha}C_{\alpha}(\mathbf{W})$ is the Hamiltonian and $Z_{\mathbf{A},\la}=\int_{\mathbb{W}_\mathbf{A}}e^{-H_{\la}(\mathbf{W})}d\mathbf{W}$ is the partition function for fixed $\mathbf{A}$. Note that we have introduced the subscript $\la$ to stress the dependence of the quantities on the Lagrange multipliers.
The explict functional form of $Q_{\la}(\mathbf{W}|\mathbf{A})$ can be obtained only by further specifying the functional form of the constraints as well.\\

\paragraph*{Parameters estimation.} 
The conditional probability distribution defined in eq. (\ref{eq:q_exprg}) depends on the vector of unknown parameters $\la$: a recipe is, thus, needed to provide their numerical estimation.
In alignment with previous results \cite{garlaschelli2008maximum,squartini2011analytical,squartini2017reconnecting}, we now extend the maximum-likelihood recipe to deal with the conditional probability distribution we are considering here. Indeed, since we do not have access to the empirical adjacency matrix $\mathbf{A}^*$, it is not possible for us to compute the usual likelihood function $Q_{\la}(\mathbf{W}^*|\mathbf{A}^*)$ as a function of the parameters $\la$.
However we can go back to the more general problem from which the usual maximization of the likelihood derives, i.e. the maximization of the constrained entropy - in our case, the conditional expression (\ref{eq:lagr}) - and obtain a corresponding generalized likelihood that requires only the available information about the network. 

Let us define $\la^*$ as the value of the parameters for which the constraints are satisfied, that is $\langle\vec{C}\rangle^*=\vec{C}^*$. By construction, the value $\la^*$ is such that the gradient of the Lagrangean $\mathcal{L}$ is zero. Importantly, in the Appendix we show that it is also the value that maximizes the \emph{generalized likelihood}

\begin{equation}\label{eq:baselike}
\mathcal{G}(\la)=-H_{\la}(\langle\mathbf{W}\rangle^*)-\sum_{\mathbf{A}\in\mathbb{A}}P(\mathbf{A})\log Z_{\mathbf{A},\la}
\end{equation}
where $\langle\mathcal{\mathbf{W}}\rangle^*$ indicates the unconditional ensemble average of $\mathbf{W}$ when the desired constraints are satisfied. The use of this notation is legitimate by the fact that throughout this study we will consider linear constraints of the form
\begin{equation}
H_{\la}(\mathbf{W})=\sum_{j\neq i}\lambda_{ij}w_{ij},
\end{equation}
so that $\langle H_{\la}\rangle=\sum_{j\neq i}\lambda_{ij}\langle w_{ij}\rangle=H_{\la}(\langle\mathbf{W}\rangle)$. The definition (\ref{eq:baselike}) is justified by the relationship between likelihood and entropy proved below. Let us focus on the expression of the conditional entropy defined in eq. (\ref{eq:condent}): using eq. (\ref{eq:q_exprg}) we can rearrange (and rename) it as

\begin{equation}
S(\la)=-\sum_{\mathbf{A}\in\mathbb{A}}P(\mathbf{A})[-\langle H_{\la}\rangle_{\mathbf{A}}-\log Z_{\mathbf{A},\la}].
\end{equation}
By evaluating $S(\la)$ in $\la^*$, we obtain
\begin{eqnarray}
S(\la^*)&=&\langle H_{\la^*}\rangle+\sum_{\mathbf{A}\in\mathbb{A}}P(\mathbf{A})\log Z_{\mathbf{A},\la^*}\nonumber\\
&=&H_{\la^*}(\langle\mathbf{W}\rangle^*)+\sum_{\mathbf{A}\in\mathbb{A}}P(\mathbf{A})\log Z_{\mathbf{A},\la^*}\nonumber\\
&=&-\mathcal{G}(\la^*).
\end{eqnarray}
Notice that the starting point of our derivation was the definition of conditional entropy involving two ensemble averages, over both sets of binary configurations and link weight assignments. 
After its evaluation in $\la^*$, the average over the set of link weight assignments has reduced to a single term, i.e. $H_{\la^*}(\langle\mathbf{W}\rangle^*)$, while the average over the space of binary configurations has survived.

\begin{figure*}[!t]
\centering
\includegraphics[width=0.49\textwidth]{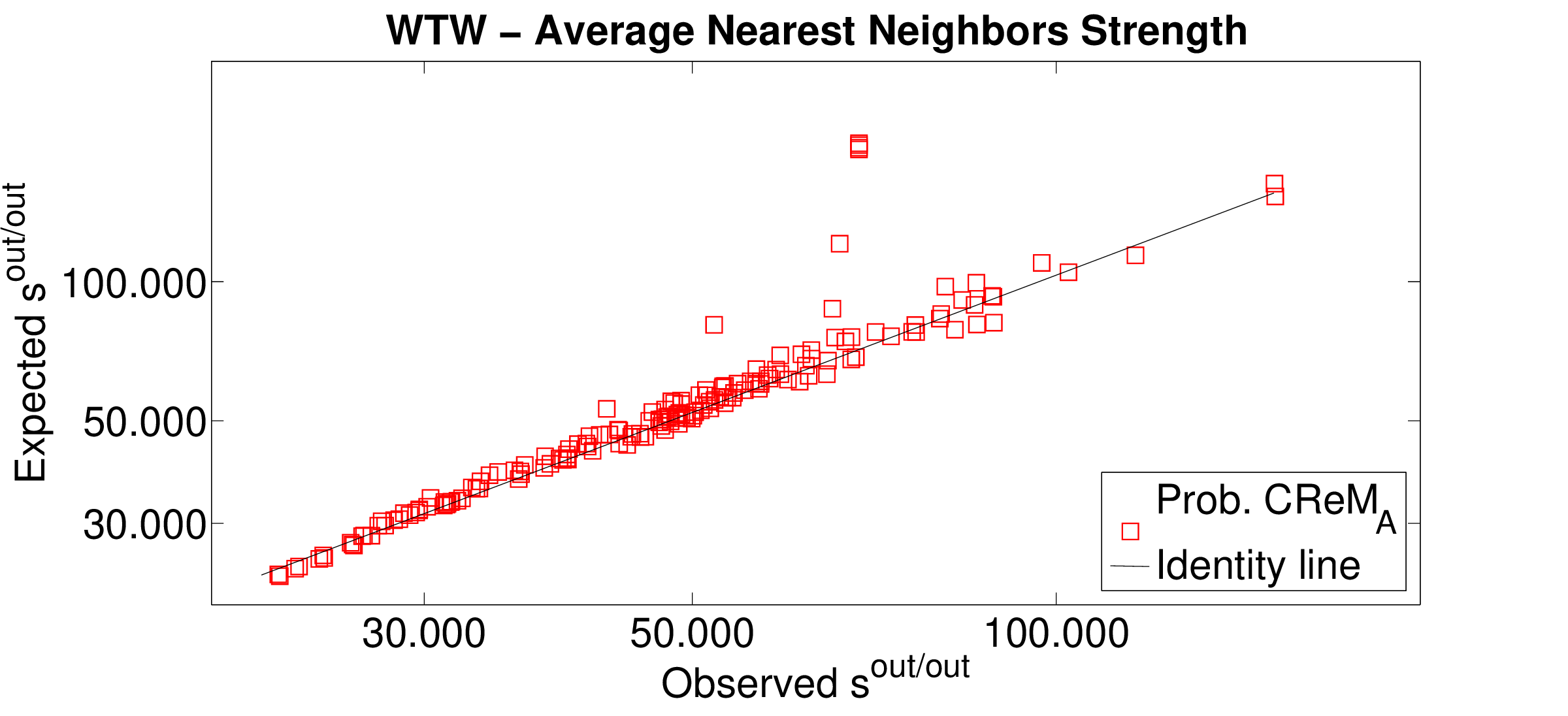}
\includegraphics[width=0.49\textwidth]{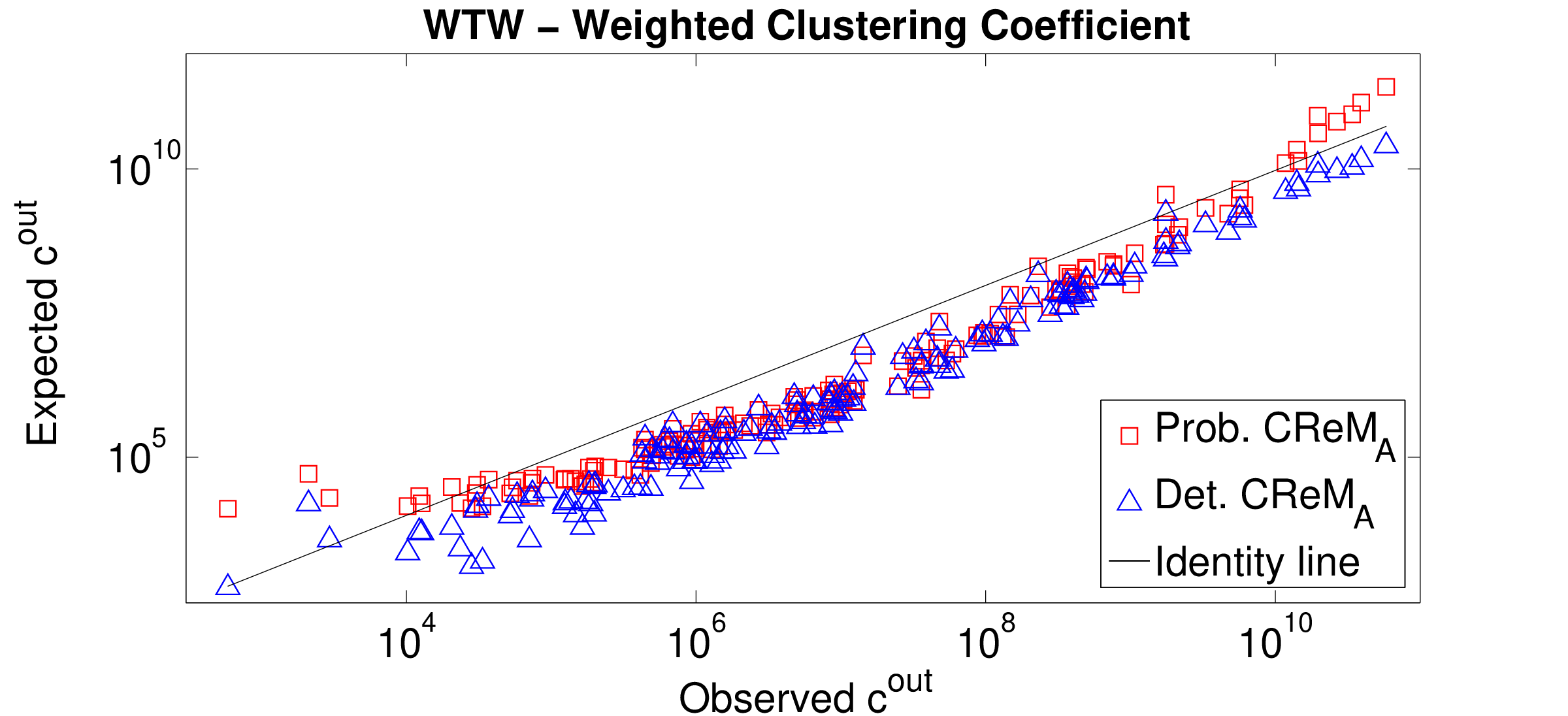}\\
\includegraphics[width=0.49\textwidth]{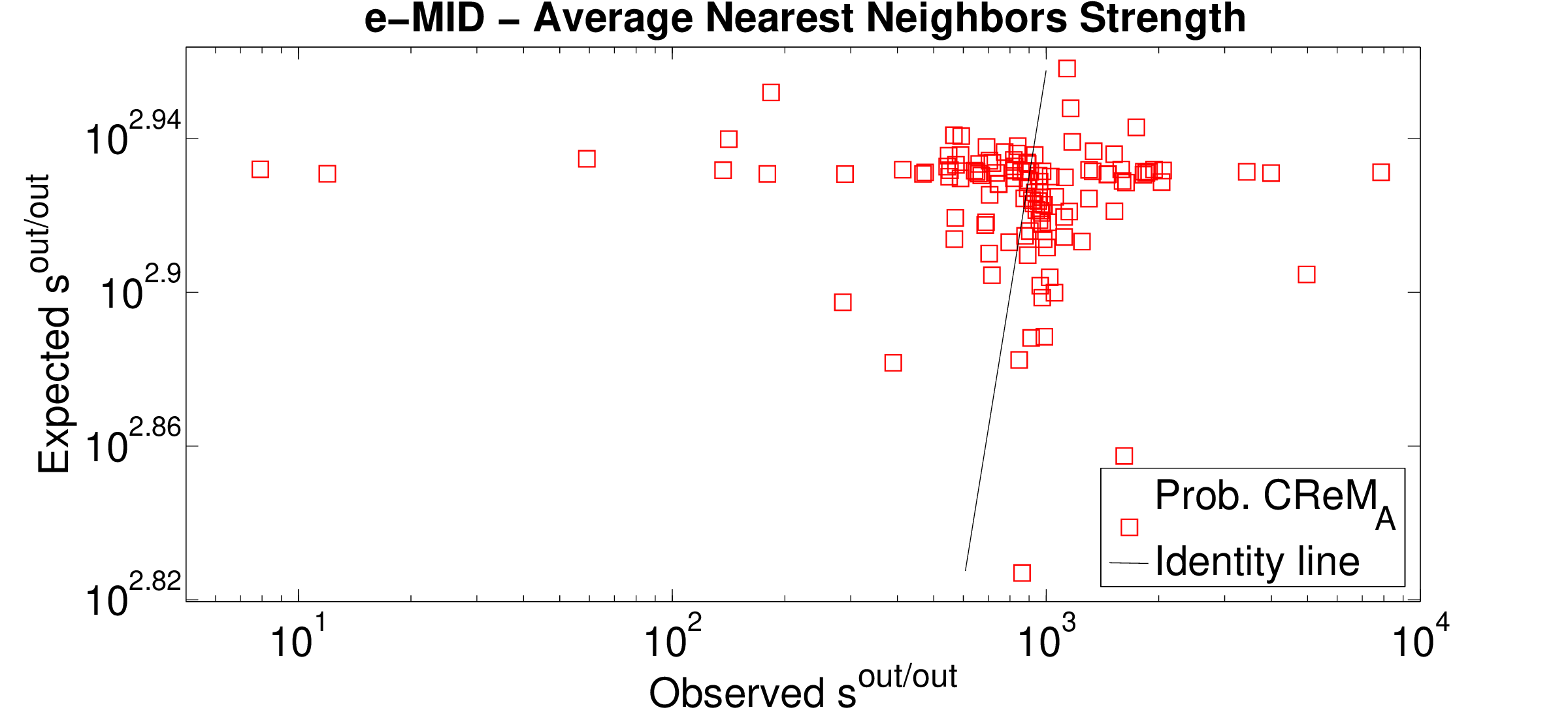}
\includegraphics[width=0.49\textwidth]{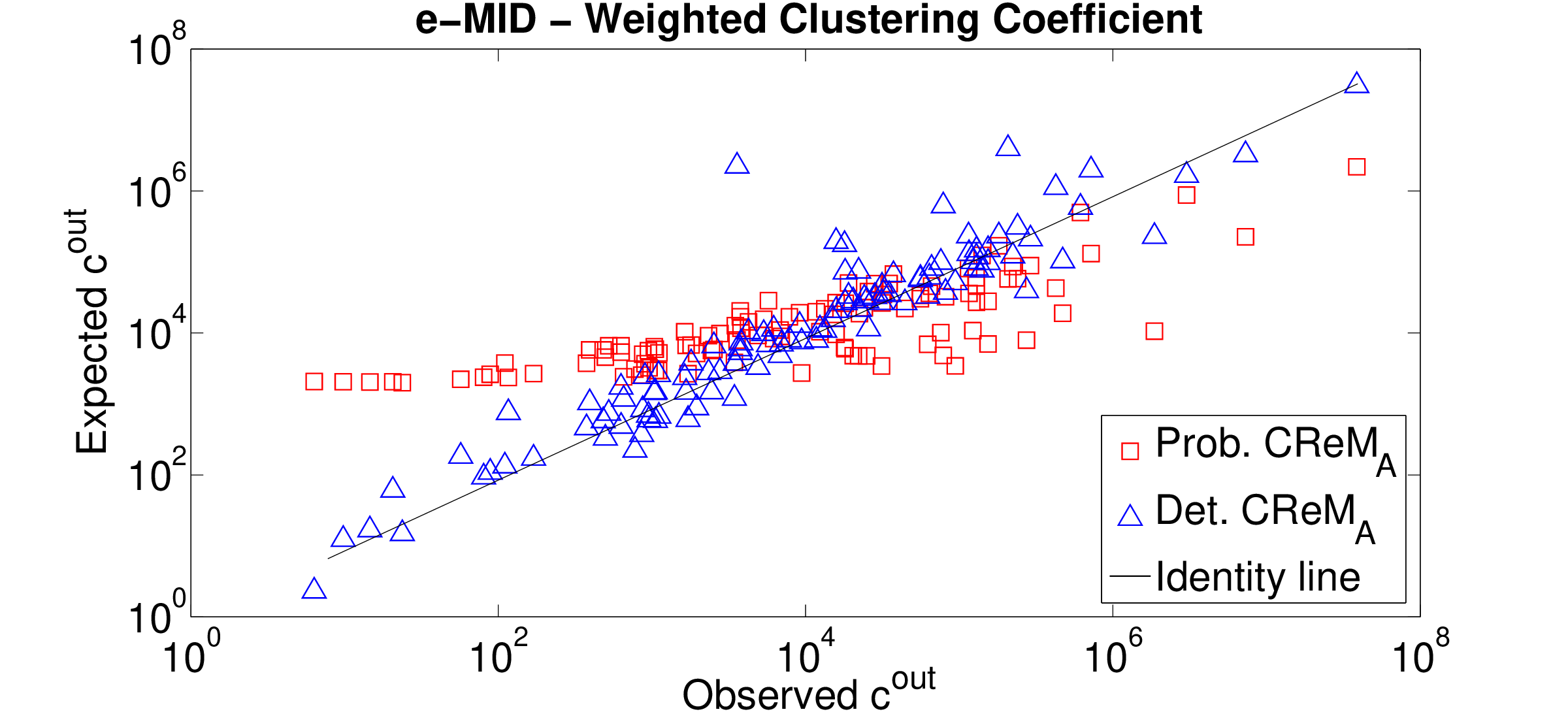}
\caption{Test of the effectiveness of the \crema\:model in reproducing the average nearest neighbors strength (left panels) and the weighted clustering coefficient (right panels) for the World Trade Web in the year 1990 (top panels) and e-MID in the year $2010$ (bottom panels). The chosen probability distributions for the binary estimation step are the one defining the density-corrected Gravity Model (red squares - see eq. \ref{eq:cimibin}) and the one defining the actual configuration (blue triangles - see eqs. \ref{eq:sys_a2}). The latter choice perfectly recovers the observed values of the ANNS that lie on the identity line (drawn as a black, solid line); the WCC is reproduced with a much higher accuracy as well.}
\label{fig1}
\end{figure*}

\subsection*{Constraining the strengths: the \crema\:model}\label{sub:str}

Let us now instantiate our CReM framework for the set of constraints usually considered (and empirically accessible) for the case of the reconstruction of financial networks, i.e. the out- and in-strength sequences, $s_i^{out}(\mathbf{W})=\sum_{j\neq i}w_{ij},\:\forall\:i$ and $s_i^{in}(\mathbf{W})=\sum_{j\neq i}w_{ji},\:\forall\:i$.
Imposing these constraints means introducing the Hamiltonian
\begin{eqnarray}
H(\mathbf{W})&=&\sum_{i=1}^N \left[\beta_i^{out}s_i^{out}(\mathbf{W})+\beta_i^{in}s^{in}_i(\mathbf{W})\right]\nonumber\\
&=&\sum_{i=1}^N\sum_{j\neq i} (\beta_i^{out}+\beta_j^{in})w_{ij},
\end{eqnarray}
which induces the partition function
\begin{eqnarray}
Z_{\mathbf{A}}&=&\prod_{i=1}^N\prod_{j\neq i}\left[\int_{0}^{\infty}e^{-(\beta_i^{out}+\beta_j^{in})w_{ij}}dw_{ij}\right]^{a_{ij}}\nonumber\\
&=&\prod_{i=1}^N\prod_{j\neq i} \left(\frac{1}{\beta_i^{out}+\beta_j^{in}}\right)^{a_{ij}}.
\end{eqnarray}
Using eq. (\ref{eq:q_exprg}), we can write \begin{equation}
Q(\mathbf{W}|\mathbf{A})=\prod_{i=1}^N\prod_{j\neq i}q_{ij}(w_{ij}|a_{ij})
\label{eq:QQ}
\end{equation}
where $q_{ij}(w=0|a_{ij}=1)=0$ and 
\begin{equation}\label{eq:condca}
q_{ij}(w|a_{ij}=1)=\left\{\begin{array}{ll}(\beta^{out}_i+\beta^{in}_j)e^{-(\beta^{out}_i+\beta^{in}_j)w}&w>0\\
0&w\le0\end{array}\right.
\end{equation}
for each positive weight $w_{ij}$, showing that each pair-specific weight distribution conditional on the existence of the link is exponential with parameter $\beta^{out}_i+\beta^{in}_j$. 

Now, in order to determine the values of the vectors of parameters $\vec{\beta}^{out}$ and $\vec{\beta}^{in}$ we maximize the generalized likelihood defined in eq. (\ref{eq:baselike}), which reads
\begin{eqnarray}\label{eq:like_A}
\mathcal{G}_{\text{CReM}_A}&=&-\sum_{i=1}^N \left[s_i^{out^*}(\mathbf{W})\beta_i^{out}+s_i^{in^*}(\mathbf{W})\beta_i^{in}\right]\nonumber\\
&&+\sum_{i=1}^N\sum_{j\neq i} f_{ij}\log (\beta^{out}_i+\beta^{in}_j)
\end{eqnarray}
where the quantity 
\begin{equation}
f_{ij}\equiv\sum_{\mathbf{A}\in\mathbb{A}}P(\mathbf{A})a_{ij}=\langle a_{ij}\rangle
\label{eq:fij}
\end{equation}
represents the expected value of $a_{ij}$ over the ensemble of binary configurations, i.e. the \emph{marginal} probability of a directed edge existing from node $i$ to node $j$ in the reconstructed binary ensemble, irrespective of whether edges are generated independently of each other by the binary reconstruction method.
This makes our formulation entirely general with respect to the binary reconstruction method taken as input, as we made no assumption on the structure of $P(\mathbf{A})$. 
In particular, the \emph{joint} probability $P(\mathbf{A})$ for all links in network $\mathbf{A}$ collectively appearing need not necessarily factorize as $P(\mathbf{A})=\prod_{i=1}^N\prod_{i\ne j}f_{ij}^{a_{ij}}(1-f_{ij})^{1-a_{ij}}$ as in models with independent edges. For instance, in the case of the microcanonical binary configuration model (defined by sharply constraining the degrees), $P(\mathbf{A})$ is the uniform distribution over all graphs with the a given degree sequence and cannot be factorized, as all links are mutually dependent through the sharp constraint. On the other hand, when the canonical binary configuration model is considered, $P(\mathbf{A})$ factorizes and $f_{ij}$ coincides with the connection probability $p_{ij}$ defining the model itself (see also the Appendix) \cite{squartini2018reconstruction}. Both variants of the binary configuration model, as well as any other binary reconstruction method, can be taken as input into our conditional reconstruction method by specifying the corresponding $P(\mathbf{A})$. Note that, in cases where the explicit expression for $P(\mathbf{A})$ is not available, one can still sample this distribution by taking multiple outputs of the binary reconstruction method and replacing averages over $P(\mathbf{A})$ with sample averages.

Now, differentiating eq. (\ref{eq:like_A}) with respect to $\beta^{out}_i$ and $\beta^{in}_i$ yields the system of $2N$ coupled equations
\begin{equation}\label{eq:sys_a1}
\left\{\begin{array}{ll}
\langle s_i^{out}\rangle&=\sum_{j\neq i}\frac{f_{ij}}{\beta^{out}_i+\beta^{in}_j}=s_i^{out^*},\:\forall\:i\\
\langle s_i^{in}\rangle&=\sum_{j\neq i}\frac{f_{ji}}{\beta^{out}_j+\beta^{in}_i}=s_i^{in^*},\:\forall\:i
\end{array}\right.
\end{equation}
where $\langle w_{ij}\rangle=\frac{f_{ij}}{\beta^{out}_i+\beta^{in}_j}$ and $f_{ij}$ is taken as given - therefore excluded from the estimation procedure.

In what follows, we consider explicitly the case where \emph{no entry} of the empirical adjacency matrix $\mathbf{A}^*$ is known, so that all entries have to be dealt with probabilistically, in line with previous research in the field.
However, an important feature of our framework is that it can incorporate the knowledge of \emph{any set} of entries of $\mathbf{A}^*$ as well. This means that, if we are certain about the presence ($a_{ij}^*=1$) or absence ($a_{ij}^*=0$) of certain edges, this deterministic knowledge will be reflected in the corresponding marginal connection probability being $f_{ij}=a_{ij}^*$. As an extreme example, let us consider the case in which \emph{all entries} are known. In this case, we are led to the maximally informative specification $f_{ij}\equiv a_{ij}^*,\:\forall\:i\neq j$ further impliying that the system to be solved becomes
\begin{equation}\label{eq:sys_a2}
\left\{\begin{array}{ll}
\langle s_i^{out}\rangle&=\sum_{j(\neq i)}\frac{a_{ij}^*}{\beta^{out}_i+\beta^{in}_j}=s_i^{out^*},\:\forall\:i\\
\langle s_i^{in}\rangle&=\sum_{j(\neq i)}\frac{a_{ji}^*}{\beta^{out}_j+\beta^{in}_i}=s_i^{in^*},\:\forall\:i.
\end{array}\right.
\end{equation}

As a final general observation before moving to specific results, we would like to stress that the framework defining the \crema\:model admits, as a particular case, the \emph{Directed Enhanced Configuration Model} (DECM), i.e. the directed version of the continuous ECM \cite{gabrielli2018grand}. For more properties of the \crema\:model, see also the Appendix. The code to run the \crema\:model is freely available at \cite{repository}.

\begin{figure*}[!t]
\centering
\includegraphics[width=0.49\textwidth]{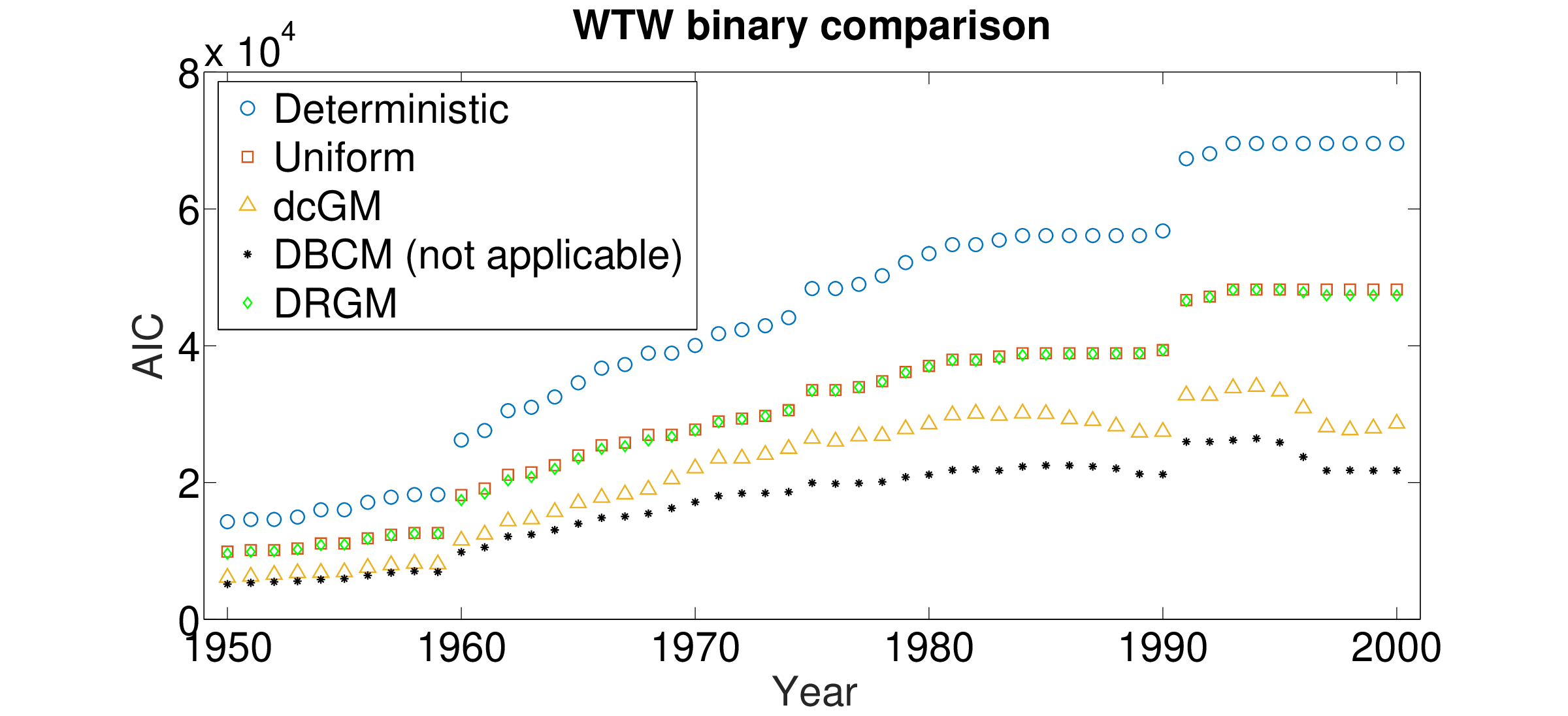}
\includegraphics[width=0.49\textwidth]{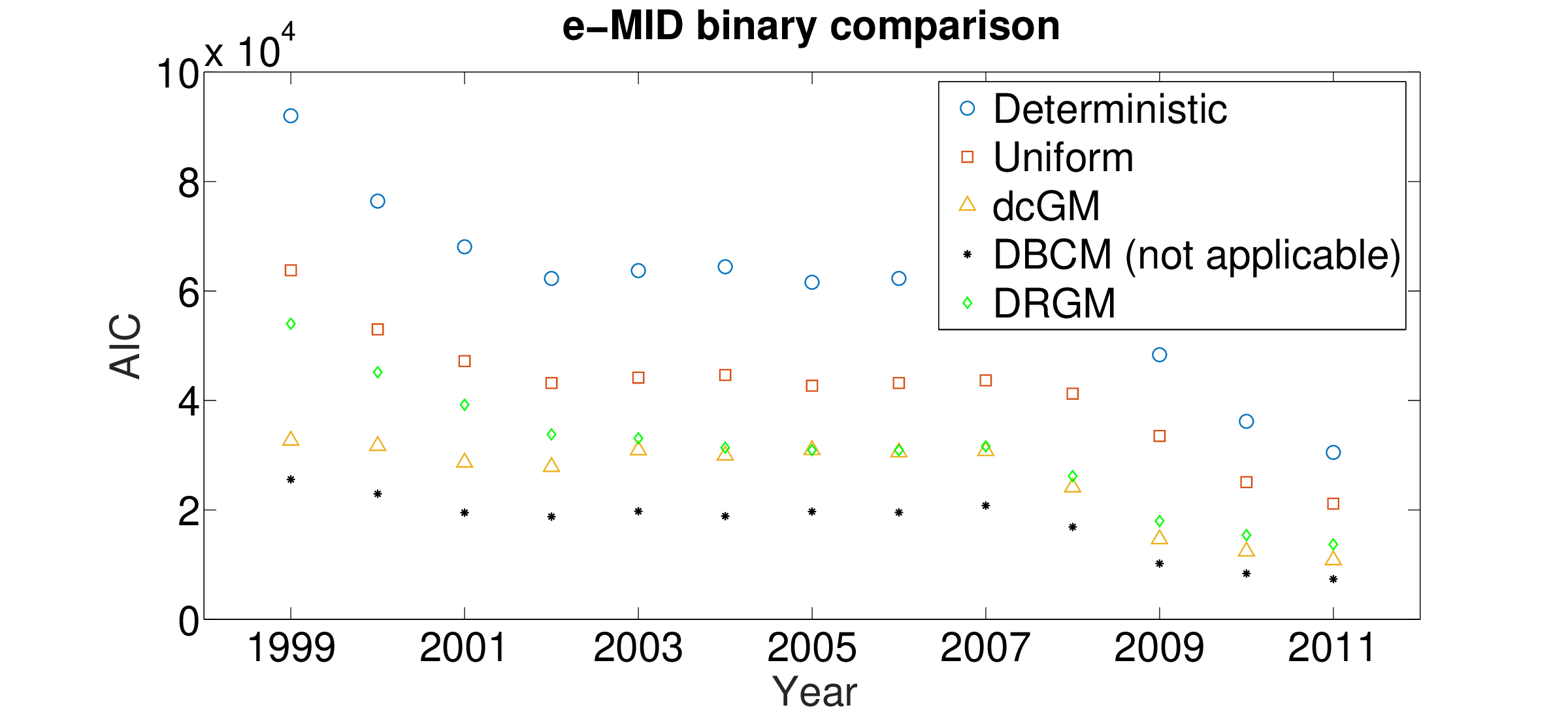}
\caption{Comparison between the binary likelihood functions for three prototypical distributions (the deterministic one, the uniform one and the dcGM one), plus the two ones induced by the popular Directed Random Graph Model (DRGM) and the Directed Binary Configuration Model (DBCM), for the WTW (across the years 1950-2000 - left panel) and e-MID (across the years 1999-2011 - right panel). As the Akaike Information Criterion certifies, the DRGM is an acceptable reconstruction model when considering very sparse networks; on the other hand, the comparison with the DBCM confirms that, \emph{in case degrees were known}, they should be preferred to the necessarily less precise fitness ansatz. Since this kind of information is practically never accessible, we need to resort to some kind of approximation: the effectiveness of the one defining the dcGM is confirmed by the evidence that the best binary \emph{applicable} model is precisely the dcGM one.}
\label{fig2}
\end{figure*}

\subsection*{Testing the \crema\:model}

Let us now explicitly test the effectiveness of the \crema\:model in reproducing two real-world systems, i.e. the World Trade Web (WTW) in the year 1990 \cite{gleditsch2002expanded} and e-MID in the year $2010$ \cite{iori2006network} (see the same references for a detailed description of the two data sets). In order to do so, we need to specify a functional form for the coefficients $\{f_{ij}\}_{i,j=1}^N$. As a first choice, let us implement the recipe

\begin{equation}\label{eq:cimibin}
f_{ij}\equiv p_{ij}^{\text{dcGM}}=\frac{zs_i^{out}s_j^{in}}{1+zs_i^{out}s_j^{in}},\:\forall\:i\neq j
\end{equation}
that defines the density-corrected Gravity Model. Upon solving the system of equations (\ref{eq:sys_a1}), we obtain the numerical value of the parameters $\vec{\beta}^{out}$ and $\vec{\beta}^{in}$ by means of which we can analytically compute the expectation of any quantity of interest (via the so-called \emph{delta method}, see also \cite{squartini2011analytical}). In particular, we have focused on (one of the four versions of) the \emph{average nearest neighbors strength} (ANNS) \cite{squartini2011randomizing}

\begin{equation}
s^{out/out}_i=\frac{\sum_ja_{ij}s_j^{out}}{k_i^{out}}
\end{equation}
and on (one of the four versions of) the \emph{weighted clustering coefficient} (WCC) \cite{squartini2011randomizing}

\begin{equation}
c^{out}_i=\frac{\sum_{j(\neq i)}\sum_{k(\neq i,j)}w_{ij}w_{jk}w_{ik}}{k_i^{out}(k_i^{out}-1)};
\end{equation}
the comparison between the observed and the expected value of the quantities above is shown in fig. \ref{fig1} for both systems. As a second choice, let us implement the deterministic recipe

\begin{equation}\label{eq:det}
f_{ij}\equiv a_{ij}^*,\:\forall\:i\neq j
\end{equation}
whose effectiveness in reproducing the ANNS and the WCC is shown in fig. \ref{fig1} as well. Notice that the WCC is now reproduced much more accurately, a result further confirming that binary constraints affects the weights estimation as well; on the other hand, the ANNS is perfectly reproduced: specifying the network topology and the weighted marginals is, in fact, enough to recover the observed values.

\begin{figure*}[t!]
\centering
\includegraphics[width=0.49\textwidth]{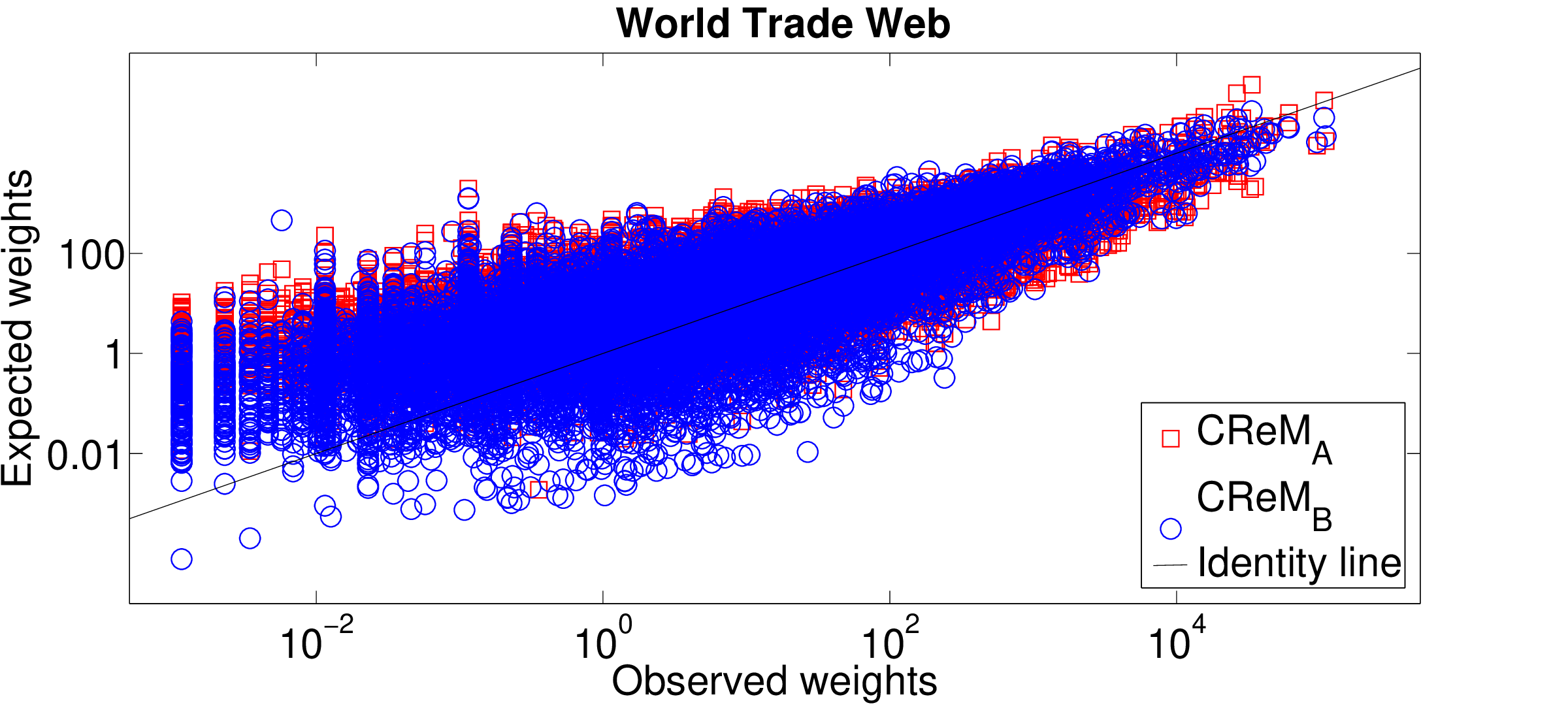}
\includegraphics[width=0.49\textwidth]{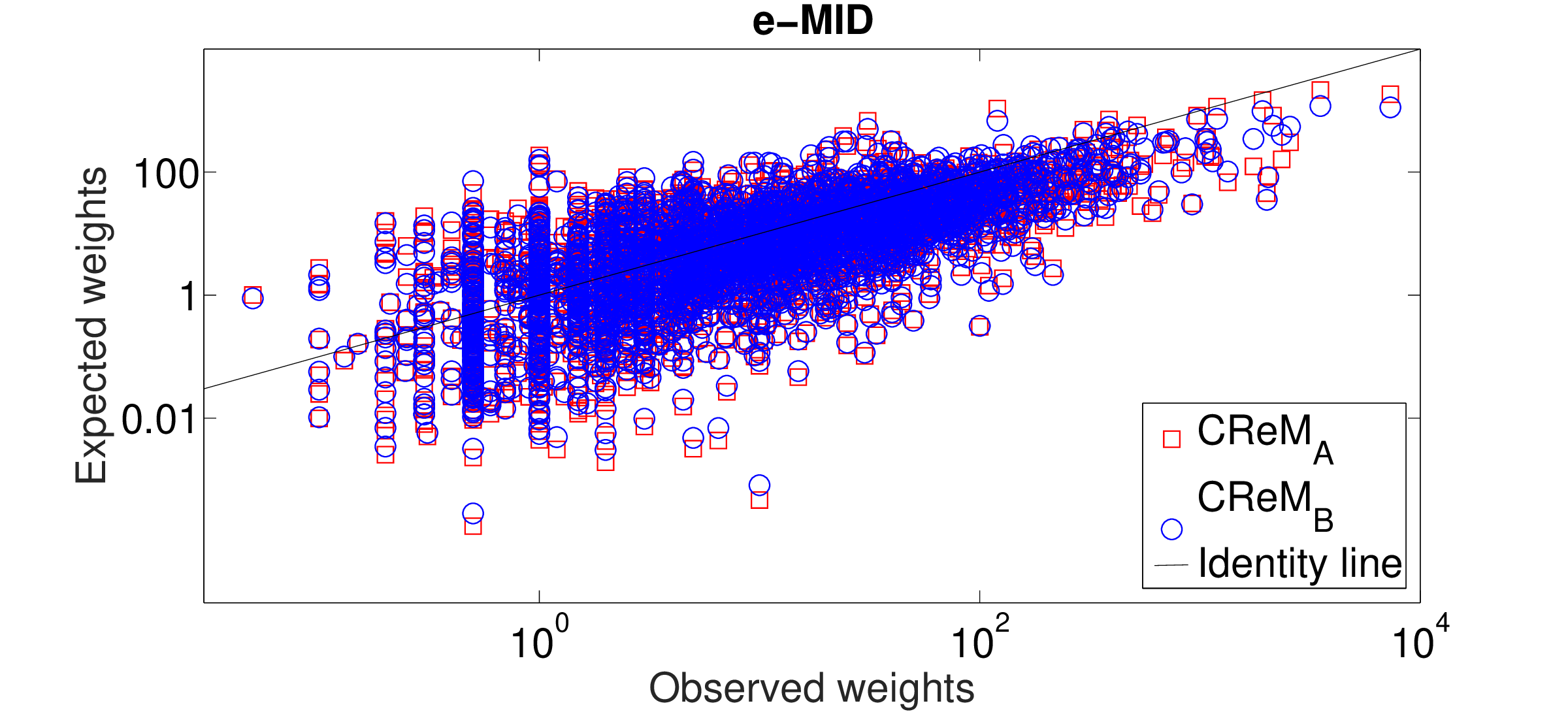}
\caption{Comparison between the realized (positive) and the corresponding expected values of the link weights for the World Trade Web in the year 1990 \cite{gleditsch2002expanded} (left panel) and for e-MID in the year 2010 \cite{iori2006network} (right panel). Two different kinds of expectations were considered: the ones coming from the \crema\:model (red squares) and the ones provided by the MaxEnt model (blue circles).
The figure shows that the expected weights of the \cremb\:model are at least as good as those of the \crema\:model, and generally even more narrowly scattered along the identity line. 
This observation is confirmed by the calculation of the Pearson correlation coefficients between realized and expected link weights: such coefficients equal $r_\text{\crema}\simeq0.6$, $r_\text{\cremb}\simeq0.75$ for the World Trade Web and $r_\text{\crema}\simeq0.44$, $r_\text{\cremb}\simeq0.5$ for e-MID.}
\label{fig3}
\end{figure*}

\subsection*{Comparing binary reconstruction methods}

As we have seen, our framework allows for any probability distribution to be taken as input to address the topology reconstruction step. We may, thus, ask what is the best recipe to reconstruct a given real-world network binary structure. In order to provide an answer, let us consider again the score function

\begin{equation}
\ln Q(\mathbf{W}^*)=\ln P(\mathbf{A}^*)+\ln Q(\mathbf{W}^*|\mathbf{A}^*)
\end{equation}
and focus on the addendum $\ln P(\mathbf{A}^*)$. Three prototypical distributions can be considered and compared:

\begin{itemize}
\item \emph{deterministic distribution}: this choice implements the maximally informative position $P(\mathbf{A})=\delta_{\mathbf{A},\mathbf{A}^*}$ (see also eq. \ref{eq:det}) and is equivalent to assuming that the empirical adjacency matrix $\mathbf{A}^*$ is known;
\item \emph{uniform probability distribution}: this choice corresponds to the maximally uninformative position $P(\mathbf{A}^*)=2^{-N(N-1)}$ (i.e. $f_{ij}\equiv\frac{1}{2},\:\forall\:i\neq j$);
\item \emph{dcGM probability distribution}: this choice implements the recipe $P(\mathbf{A}^*)=\prod_i\prod_{j\neq i}f_{ij}^{a_{ij}^*}(1-f_{ij})^{1-a_{ij}^*}$, with $f_{ij}\equiv p_{ij}^{\text{dcGM}}=\frac{zs_i^{out}s_j^{in}}{1+zs_i^{out}s_j^{in}},\:\forall\:i\neq j$ defining the density-corrected Gravity Model.
\end{itemize}

In order to test the performance of the three competing models above, let us invoke the Akaike Information Criterion (AIC) \cite{akaike1974new} to select the model with the best trade-off between accuracy and parsimony. The AIC value is defined as
\begin{equation}
\mbox{AIC}_m=2k_m-2\mathcal{L}_m
\end{equation}
for each model $m$ in the basket, with $\mathcal{L}_m$ indicating the log-likelihood value of model $m$ and $k_m$ indicating the number of parameters defining it (and to be estimated\footnote{Note that, since we assume from the beginning that the in- and out-strengths of all nodes are known independently of the model used, we do not count the strengths among the model parameters.}). For the three alternatives above we have that

\begin{eqnarray}
\mbox{AIC}_{\text{deterministic}}&=&2N(N-1),\\
\mbox{AIC}_{\text{uniform}}&=&2N(N-1)\ln2,\\
\mbox{AIC}_{\text{dcGM}}&=&2(1-\mathcal{L}_{\text{dcGM}})
\end{eqnarray}
where we have used the fact that the uniform model is non-parametric (i.e. $k_\text{uniform}=0$) and is characterized by the log-likelihood $\mathcal{L}_{\text{uniform}}=-N(N-1)\ln2$, while the deterministic model is characterized by a probability $P(\mathbf{A}^*)=1$ (implying $\mathcal{L}_{\text{deterministic}}=0$) and a number of parameters equal to $k_{\text{deterministic}}=N(N-1)$ (all off-diagonal entries of $\mathbf{A}^*$ are separately specified). Moreover, we have added the comparison with the Directed Random Graph Model and the Directed Binary Configuration Model, respectively defined by

\begin{eqnarray}
\mbox{AIC}_\text{RGM}&=&2(1-\mathcal{L}_\text{DRGM}),\\
\mbox{AIC}_\text{DBCM}&=&2(2N-\mathcal{L}_\text{DBCM})
\end{eqnarray}
where $\mathcal{L}_{\text{DRGM}}=L\ln p+(N(N-1)-L)\ln(1-p)$, with $p=\frac{L}{N(N-1)}$ and $\mathcal{L}_{\text{DBCM}}=\sum_i\sum_{j(\neq i)}a_{ij}\ln p_{ij}^\text{DBCM}+(1-a_{ij})\ln(1-p_{ij}^\text{DBCM})$ with $p_{ij}^\text{DBCM}=\frac{x_iy_j}{1+x_iy_j},\:\forall\:i\neq j$ \cite{squartini2018reconstruction}. The criterion prescribes to prefer the model whose AIC value is minimum. Upon looking at fig. \ref{fig2}, one realizes that the DRGM is a poorly-performing reconstruction model when the network link density is close to $0.5$ as its performance cannot be distinguished from the one of the uniform model. When considering very sparse networks, on the other hand, knowing the link density means adding a non-trivial piece of information, potentially reducing the uncertainty about a given network structure to a large extent: this seems indeed to be the case for several temporal snapshots of eMID. On the other hand, the comparison with the DBCM confirms that, in case degrees were known, they should be preferred to the necessarily less-precise fitness ansatz: however, as they are never known, including the DBCM is a merely academic exercise\footnote{In this case, in fact, one would have enough information to run the DECM - a model that we have explicitly excluded from our comparison since completely unrealistic.}. Nevertheless, the previous analysis still conveys an important message: the structure of real-world networks is characterized by a large amount of redundancy, as evident by noticing that the AIC value of the DBCM is much lower than the AIC of the fully deterministic model. Hence, the structure of complex networks can indeed by explained by only constraining a bunch of statistics, as exemplified by the degrees: however, since this kind of information is practically never accessible, we need to resort to some kind of approximation - whence our definition of the dcGM.

The importance of employing a method able to provide a reliable estimate of a network topology becomes evident when considering the problem of quantifying systemic risk (see \cite{squartini2018reconstruction} and references therein). To this aim, let us consider the \emph{triangular loops} arising from various `risky' triadic motifs connected to the underestimation of counterparty risk due to over-the-counter linkages in interbank networks \cite{earlywarning}. An aggregate measure of incidence of such patterns is quantified by

\begin{equation}
\overline{w}_\triangle=\frac{\sum_i\sum_{j(\neq i)}\sum_{k(\neq i,j)}w_{ij}w_{jk}w_{ki}}{\sum_i\sum_{j(\neq i)}\sum_{k(\neq i,j)}a_{ij}a_{jk}a_{ki}}=\frac{w_\triangle}{N_\triangle}
\end{equation}
i.e. the average weight per loop. Notice that the expected value of such a quantity calls for the estimation of the probability that nodes $i$, $j$ and $k$ establish a connection. For the sake of illustration, let us discuss the application of either the MaxEnt method or the Minimum-Density method to provide such an estimation. As previously discussed, the fully-connected topology output by the ME leads to $N_\triangle\simeq N^3$, i.e. to overestimating the number of cycles, in turn leading to an underestimation of systemic risk; on the other hand, the very sparse topology output by the MD leads to $N_\triangle\simeq O(1)$, i.e. to underestimating the number of cycles, in turn leading to an overestimation of systemic risk.

\begin{figure*}[!t]
\centering
\includegraphics[width=0.49\textwidth]{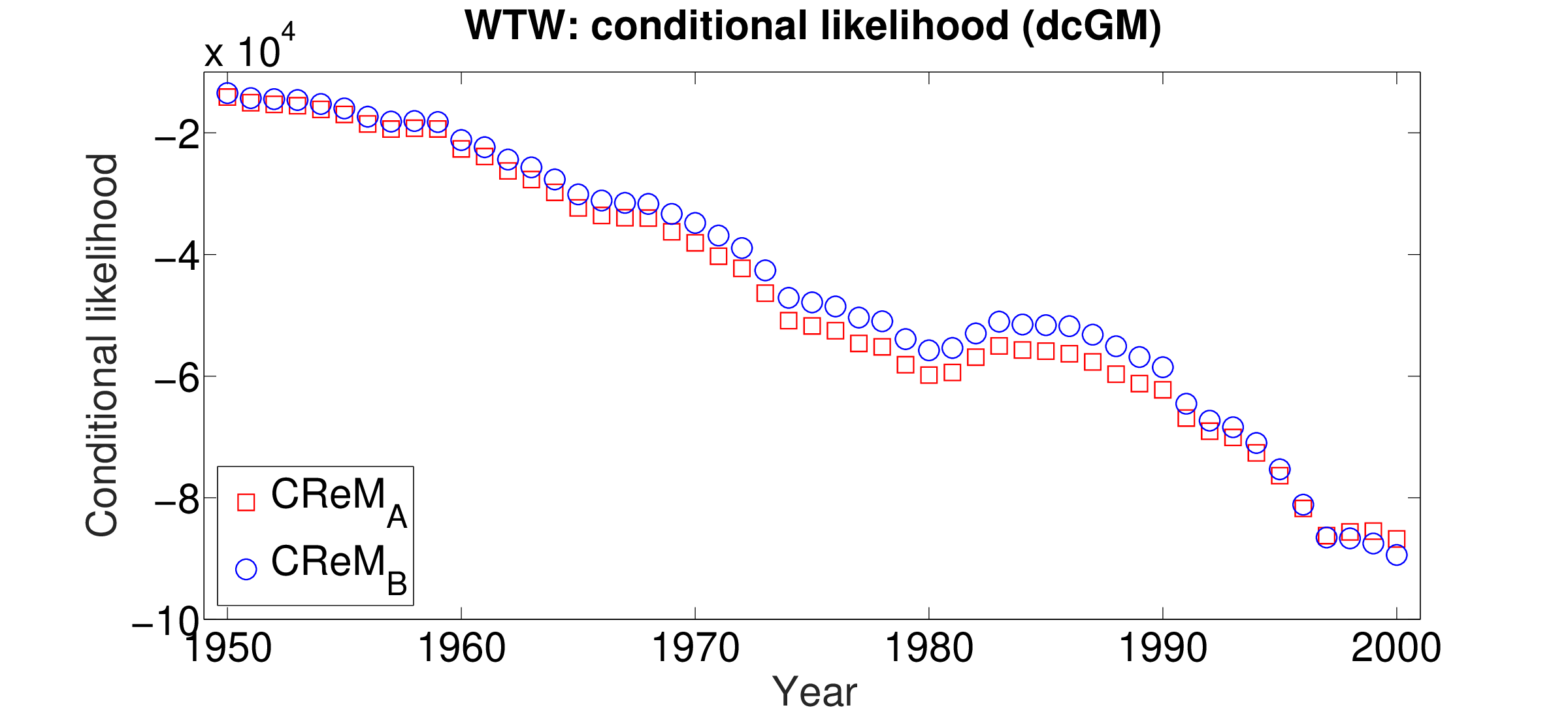}
\includegraphics[width=0.49\textwidth]{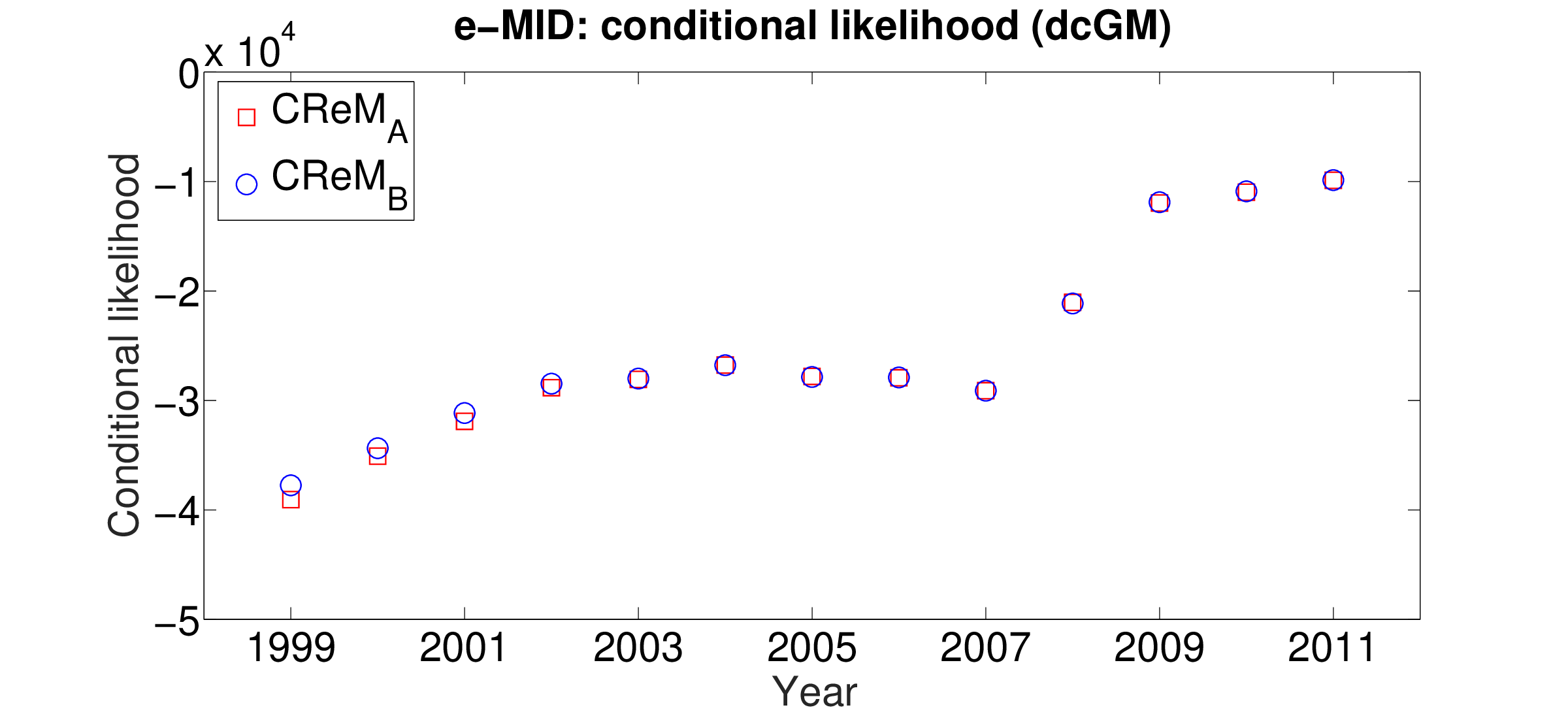}\\
\includegraphics[width=0.49\textwidth]{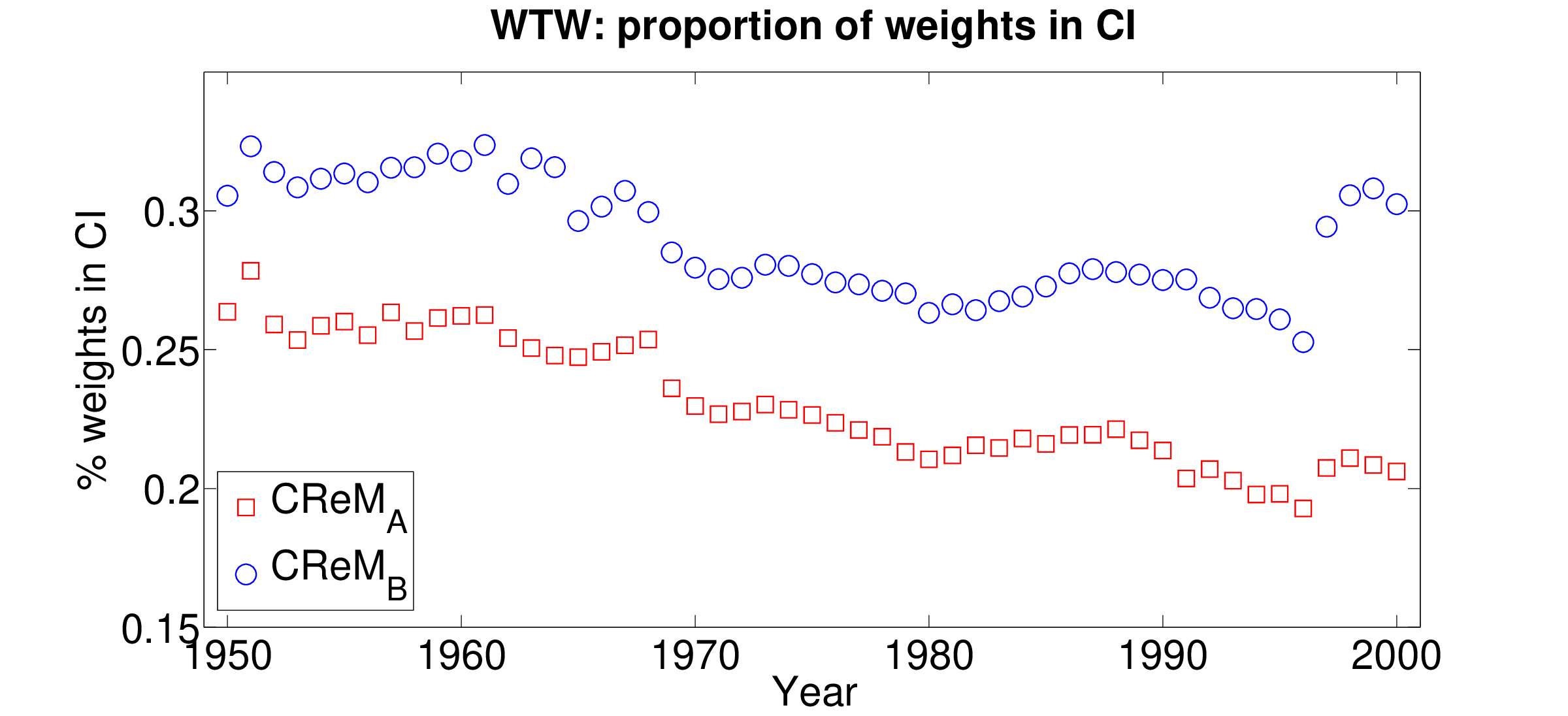}
\includegraphics[width=0.49\textwidth]{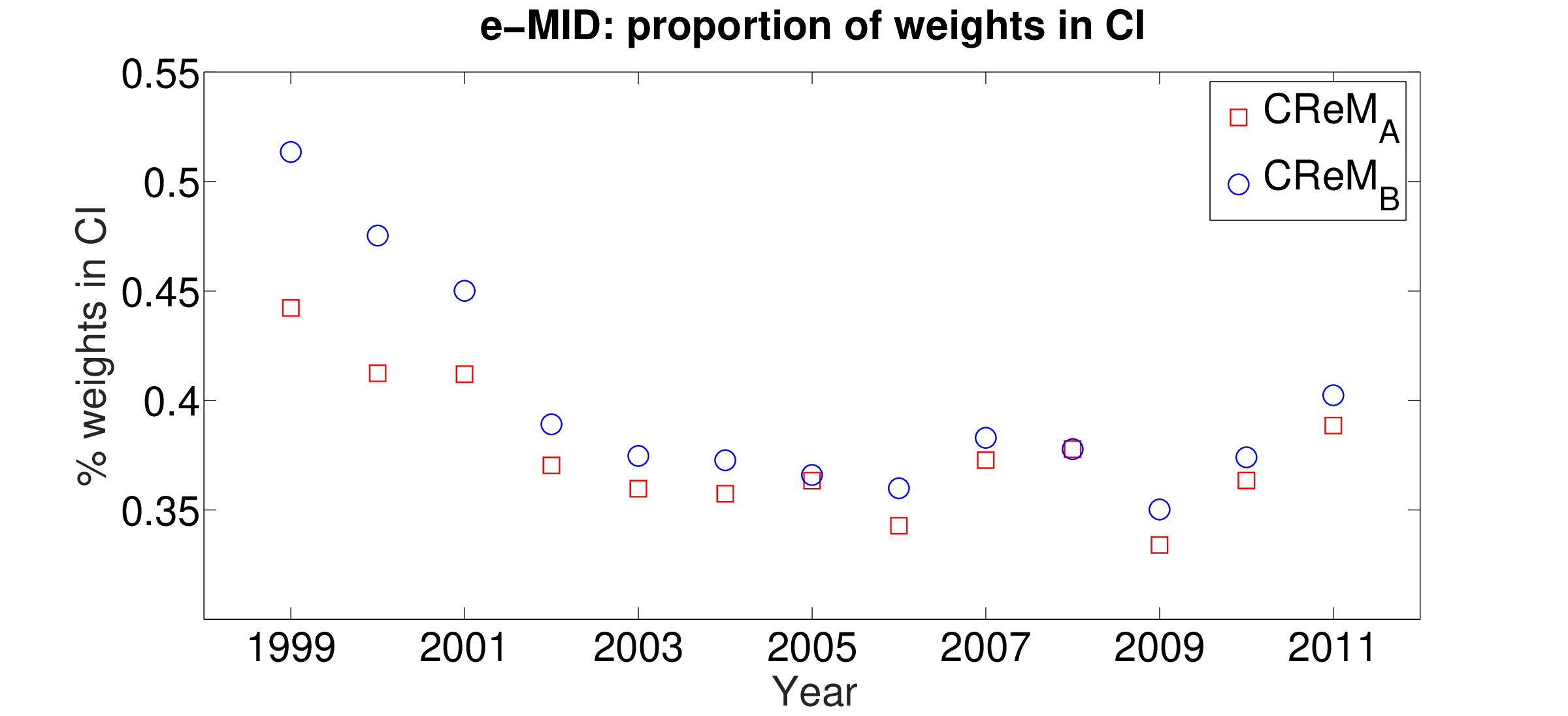}\\
\includegraphics[width=0.49\textwidth]{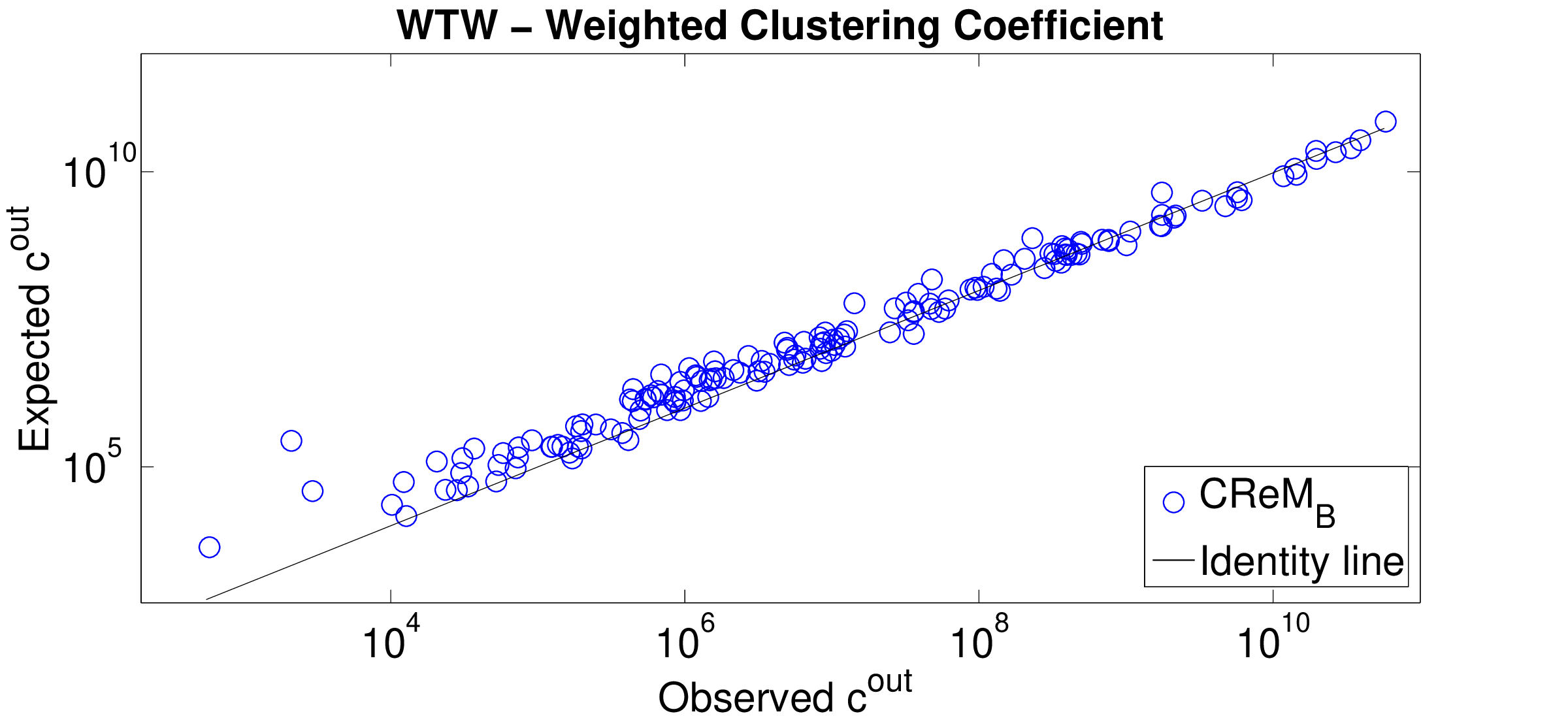}
\includegraphics[width=0.49\textwidth]{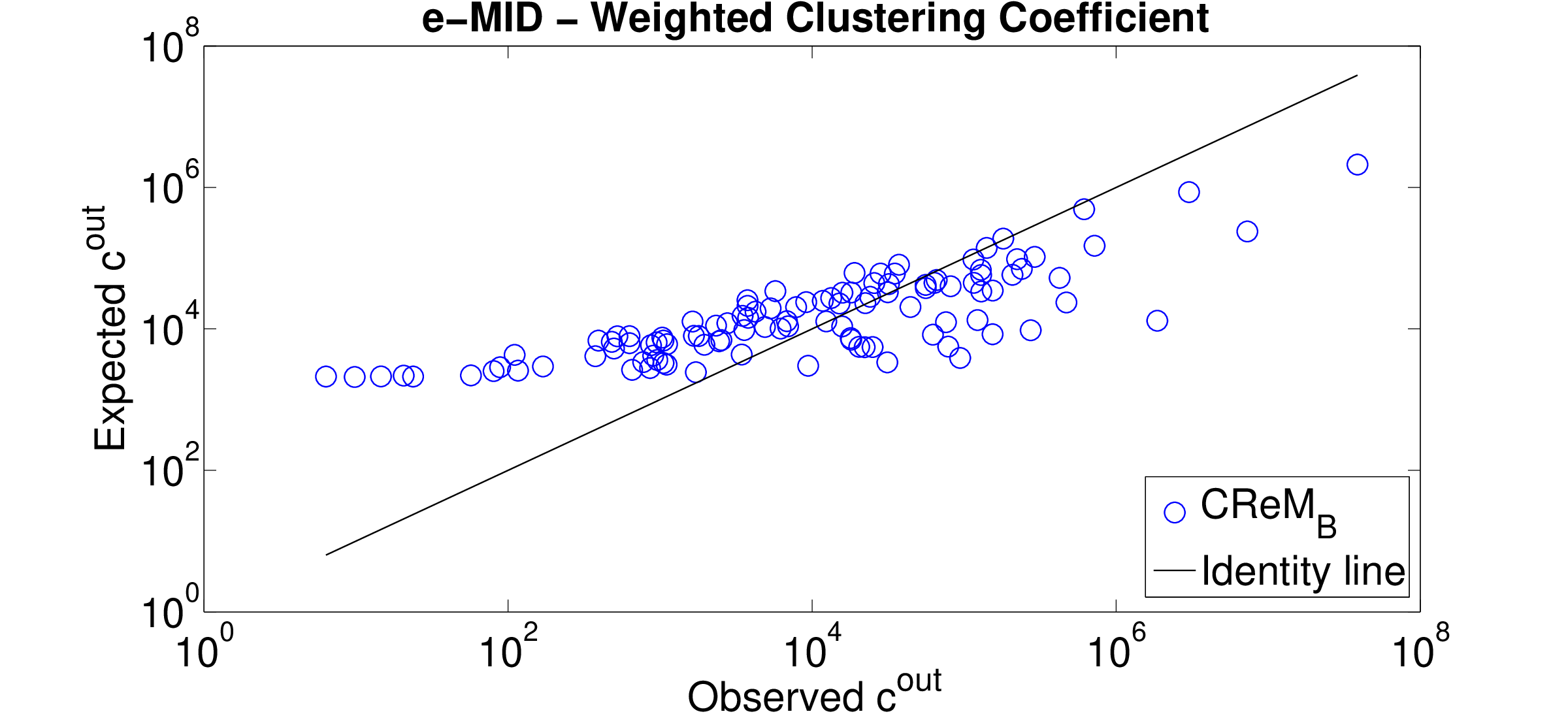}
\caption{Top panels: comparison between the conditional likelihood functions of the \crema\:and the \cremb\:models (red squares and blue circles, respectively), for the WTW (across the years 1950-2000 - left panel) and e-MID (across the years 1999-2011 - right panel). The reconstruction accuracy obtainable by employing the \cremb\:model is comparable with the one obtainable by employing the \crema\:model; still, it is achievable with much less computational effort. Middle panels: percentage of observed weights that fall into the confidence interval surrounding their estimate, for the WTW (left panel) and e-MID (right panel), in correspondence of the values $q^+=q^-=0.25$. Bottom panels: performance of the \cremb\:model in reproducing the WCC, confirming that the precision achievable by running the latter is larger than/equal to the one achievable by running the \crema\:model (analogous results hold true for the ANNS). Identity is drawn as a black, solid line.}
\label{fig4}
\end{figure*}

\subsection*{Further structuring the model:\\the \cremb\:model}\label{sub:weight}

In the previous sections we have introduced a novel framework for network reconstruction that has led to the definition of the \crema\:model. Although this model provides an accurate reconstruction of real-world economic and financial networks, its implementation still requires the resolution of $2N$ coupled non-linear equations. Moreover, as it makes the maximally random hypothesis about link weights, given the empirical in-strength and out-strength of all nodes, the model does not allow to incorporate any further ansatz or assumption about the empirical relationship between link weights and the node strenghts themselves. We now ask ourselves if it is possible to simplify the computational requirements of the \crema\:model, while more flexibily constraining its randomness (again via entropy maximization) around a structured relationship that captures some empirical regularity relating link weights to the node strengths, thus improving the accuracy of the reconstruction of the weighted network as a whole.

To this aim, let us now specify a model potentially constraining the whole set of expected weights to given values $\{w^*_{ij}\}$, where in this case the asterisk denotes the `target' value, which is not necessarily an observable one.  In order to do so, let us formally constrain the unconditional expected values of all link weights, i.e. consider a Hamiltonian reading $H(\mathbf{W})=\sum_{j\neq i}\beta_{ij}w_{ij}$. The derivation is analogous to the previous case and leads to the expression $Q(\mathbf{W}|\mathbf{A})=\prod_{j\neq i}q_{ij}(w_{ij}|a_{ij})$ with $q_{ij}(w_{ij}=0|a_{ij}=1)=0$ and

\begin{equation}\label{eq:condcb}
q_{ij}(w_{ij}|a_{ij}=1)=\beta_{ij}e^{-\beta_{ij}w_{ij}},\:w_{ij}>0
\end{equation}
i.e. to a conditional pair-specific weight distribution, $q_{ij}(w_{ij}|a_{ij}=1)$, that is exponential with parameter $\beta_{ij}$. Analogously, the generalized likelihood function can be expressed as

\begin{equation}
\mathcal{G}_{\text{CReM}_B}=-\sum_{j\neq i}w_{ij}^*\beta_{ij}+\sum_{j\neq i}f_{ij}\log\beta_{ij}
\end{equation}
and differentiating it with respect to $\beta_{ij}$ leads to the equations

\begin{equation}\label{eq:wij_b}
\langle w_{ij}\rangle=\frac{f_{ij}}{\beta_{ij}}=w_{ij}^*,\:\forall\:i\neq j
\end{equation}
that define the \cremb\:model.

Actual weights, however, can rarely be observed: hence, in order to implement the \cremb\: model, we need to replace $\{w_{ij}^*\}_{i,j=1}^N$ with a set of accessible quantities. To this aim, we look for an additional ansatz based on empirical regularities relating link weights to node strengths in the data. 
In particular, as we already mentioned we notice that the MaxEnt model introduced in eq. \eqref{eq:cw_gm} provides good estimates of the realized (i.e. positive) link weights (despite the impossibility of generating zero link weights).
Figure \ref{fig3} shows the comparison between the observed, positive weights of the World Trade Web (WTW) in the year 1990 \cite{gleditsch2002expanded} and e-MID in the year $2010$ \cite{iori2006network} and two expectations: the ones coming from the \crema\:model and the ones coming from the MaxEnt model of eq. \eqref{eq:cw_gm}.
One can see that the MaxEnt model produces expected weights that are more narrowly scattered around the empirical ones than the \crema\:model. The calculation of the Pearson correlation coefficient between the empirical and expected weights from the two models confirms that the estimates coming from the MaxEnt model show a better agreement with the data (see caption of fig. \ref{fig3}), throughout the considered time intervals.

The reason for the improved estimate in the MaxEnt model comes from the fact that the \crema\: model makes the maximally random hypothesis about link weights, based on the empirical values of the in- and out-strenghts of nodes.
Real data turn out to be more structured than this completely random expectation, the MaxEnt model better capturing the structured relation. 
At the same time, while the original MaxEnt model would assume the same positive expression \eqref{eq:cw_gm} for all link weights, the generalized framework used here allows us to embed the MaxEnt estimate into a conditional expectation for the link weight, given that the link is realized with the marginal probability $f_{ij}$ implied by the desired prior distribution $P(\mathbf{A})$. 
This is easily done by replacing the set of target expected weights $\{w_{ij}^*\}_{i,j=1}^N$ with the MaxEnt ansatz $\{\hat{w}_{ij}^{\text{ME}}\}_{i,j=1}^N$ given by eq. \eqref{eq:cw_gm} and inverting eq. \eqref{eq:wij_b} to find the corresponding tensor of coefficients $\bm{\beta}$.
This yields
\begin{equation}\label{eq:b_beta}
\beta_{ij}=\frac{f_{ij}}{\hat{w}^{\text{ME}}_{ij}}=\frac{Wf_{ij}}{s^{out}_i s^{in}_j},\:\forall\:i\neq j.
\end{equation}
Notice that this choice only requires, as input, the out- and in-strength sequences of the actual network: as a consequence, the sufficient statistics for the \crema\:and \cremb\:models coincide. Notice also that implementing the \cremb\:model requires the resolution of $O(N^2)$ \emph{decoupled} equations.

Although the choice leading to eq. (\ref{eq:b_beta}) guarantees that non-negative strengths are preserved only in case $f_{ij}>0,\:\forall\:i\neq j$, in principle, one can set $w_{ij}^*$ equal to the outcome of any other deterministic model for the link weights (e.g. IPF), not only the MaxEnt one. This would relax the requirements about the connection probability between nodes - hence allowing for zero-probability links as well - and `dress' the chosen model with a weight distribution centered around the same value generated by the deterministic implementation (thereby turning the deterministic model into a probabilistic one). The code to run the version of the \cremb\:model discussed here is freely available at \cite{repository}.

For instance, we may use a more refined recipe improving the MaxEnt ansatz to higher order. To explain this point, we need to emphasize that the MaxEnt ansatz  $w_{ij}^*=\hat{w}_{ij}^{\text{ME}}=s^{out^*}_{i}s^{in^*}_j/W^*$ introduced in eq. \eqref{eq:cw_gm} has a disadvantage: it replicates the in- and out-strengths only if a self-loop with intensity $\hat{w}_{ii}^{\text{ME}}=s^{out^*}_{i}s^{in^*}_i/W^*$ is added to each node $i$. This is easy to see by summing $\hat{w}_{ij}^{\text{ME}}$ over $i$ or $j$ to produce the resulting $s^{in^*}_j$ or $s^{out^*}_i$ respectively. In order to avoid adding self-loops, one may iteratively `redistribute' the weight $s^{out^*}_{i}s^{in^*}_i/W^*$ to all the other links. This generates a sequence of improved weights $w_{ij}^*=\hat{w}^{\text{ME}}_{ij}+\hat{w}_{ij}^{(l)}$ for any desired order $l$ of approximation \cite{density}. To this aim, at least two different recipes can be devised. The first one prescribes to redistribute the terms $s^{out^*}_{i}s^{in^*}_i/W^*$ \textit{on a complete graph with no self-loops} via the IPF algorithm. In this way, margins are correctly reproduced in the limit $l\to\infty$, with the improved weights reading $w_{ij}^*=\hat{w}^{{ME}}_{ij}+\hat{w}_{ij}^{(\infty)}$: $\hat{w}_{ij}^{(\infty)}$ can be estimated numerically, according to the iterative recipe described in [1,35]; although the final result of this procedure achieves a refined match to the enforced margins, it makes the model no longer under complete analytical control. The second one prescribes to redistribute the terms $s^{out^*}_{i}s^{in^*}_i/W^*$ \textit{on a fully connected matrix} via the IPF algorithm, discard the diagonal terms and redistribute the latter ones in an iterative fashion; in this way, the correction term is always under analytical control even if this second variant requires the explicit generation of self loops to ensure that margins are reproduced at each iteration step: for example, the full prescription of the second method, at the second iteration, reads

\begin{eqnarray}
w_{ij}^*=\begin{cases}
\hat{w}_{ij}^\text{ME}+\hat{w}_{ij}^{(2)}& \forall\:i\neq j\\
\hat{w}_{ij}^{(2)} & \forall\:i=j
\end{cases}
\end{eqnarray}
where $\hat{w}_{ij}^{(2)}=s^{out^*}_{i}s^{in^*}_is^{out^*}_{j}s^{in^*}_j/(W^*\sum_ks^{out^*}_{k}s^{in^*}_k)$. It is therefore up to the researcher to make the optimal choice between a more accurate and a more explicit version of the method, depending on the situation. Since the IPF algorithm \emph{cannot} univocally determine a way to redistribute weights (as we have seen, the answer provided by the IPF algorithm depends on how one chooses to decompose the constraints) here we have decided to use the more explicit recipe $w_{ij}^*=\hat{w}^{\text{ME}}_{ij}$, provided its agreement with the empirical weights.

Let us now compare the effectiveness of the \crema\:and the \cremb\:models in reproducing the two systems under consideration. In order to carry out the most general comparison possible, let us consider again our likelihood-based score function and focus on the second term, i.e. the proper \emph{conditional likelihood}
\begin{equation}\label{eq:cond}
\mathcal{G}(\la)=\ln Q(\mathbf{W}^*|\mathbf{A}^*);
\end{equation}
we have employed the symbol $\mathcal{G}$ since the expression of the conditional likelihood can be recovered by specifying the binary probability distribution $P(\mathbf{A})=\delta_{\mathbf{A},\mathbf{A}^*}$ in the expression of the generalized likelihood, i.e. eq. (\ref{eq:baselike}). In this case, $\mathcal{G}(\la)$ quantifies the effectiveness of a given model in reproducing the weighted structure of a network \emph{given} its topology.

The performance of the \crema\:and the \cremb\:models is, then, evaluated by comparing their conditional likelihood numerical values. The latter depend, respectively, on the parameters $\bm{\beta}_{\text{CReM}_A}$ and $\bm{\beta}_{\text{CReM}_B}$; thus, in order to compare our two models, we first solve eqs. (\ref{eq:sys_a1}) and (\ref{eq:b_beta}) (with $f_{ij}\equiv p_{ij}^{\text{dcGM}},\:\forall\:i\neq j$, to capitalize on the result of the comparison between binary reconstruction algorithms) and then substitute $\bm{\beta}_{\text{CReM}_A}^*$ and $\bm{\beta}_{\text{CReM}_B}^*$ back in eq. (\ref{eq:cond}). We also explicitly notice that the sufficient statistics for the \crema\:and the \cremb\:models coincide (they are, in fact, represented by the vectors of out- and in-strengths): hence, the AIC test would yield the same ranking as the one obtained by just comparing the likelihood functions.

Results are shown in fig. \ref{fig4} that confirms what we expected from the \cremb\:model, i.e. a reconstruction accuracy that is comparable with the one of the \crema\:model but still achievable with much less computational effort (in the case of the WTW, however, the even better agreement obtainable by running the \cremb\:model can be clearly appreciated).

Let us now compare the \crema\:and \cremb\:models by calculating the percentage of real weights that fall into the confidence intervals surrounding their estimates, by employing the same $q^-$ and $q^+$ values (see fig. \ref{fig4} and the Appendix for the details of the calculations): the \cremb\:model outperforms the \crema\:model in providing reliable estimates of actual weights. Notice that although the discrete versions of both the ECM and the DECM can provide error estimates, their computation is much easier within the novel, continuous framework considered here.

\section{Discussion}\label{sec:4}

The extension of the Exponential Random Graph framework to account for conditional probability distributions aims at filling a methodological gap: defining a recipe for \emph{unbiased weight assessment}, currently missing within the class of network reconstruction models.

The vast majority of the algorithms that have been proposed so far, in fact, combine methodologically different steps to estimate the purely topological network structure and the link weights, potentially distorting the entire procedure: as the derivation of our Conditional Reconstruction Method proves, the topological information (summed up by the set of coefficients $\{f_{ij}\}_{i,j=1}^N$) affects the estimation of link weights as well - see eqs. (\ref{eq:sys_a1}), (\ref{eq:sys_a2}), (\ref{eq:wij_b}), etc.

These observations point out that a first source of bias is encountered whenever a \emph{probabilistic} recipe for topological reconstruction is forced to output a \emph{single} outcome instead of considering the entire ensemble of admissible configurations. Indeed, (mis)using a probabilistic method by implementing it as a deterministic one leads to an (arbitrary) privilege for a single configuration instead of correctly accounting for the entire support of the probability distribution defining the method itself. Since the expectation of any quantity of interest should be taken over the entire set of admissible configurations, privileging a particular realized topology will, in general, lead to a wrong estimate of the inspected properties. Such an ``extreme'' choice is allowed only when the number of admissible configurations indeed reduces to one, i.e. only in the limiting case in which the network topology is known exactly (i.e. $f_{ij}=a_{ij}^*,\:\forall\:i\neq j$).

A second source of bias is encountered when link weights are \emph{deterministically} imposed via a recipe like the IPF algorithm, again because of the non-maximum-entropy nature of any deterministic algorithm. As a result, even in the extreme case in which \emph{all} links of a given real-world network are known \emph{ab initio}, the probability density of reproducing the weighted network with IPF-assigned weights would still be zero. By contrast, our calculations show that the correct procedure when all the topology is known is to assign weights probabilistically using eq.~\eqref{eq:condca}, with parameters fixed by eq.~\eqref{eq:sys_a2}.

Our framework overcomes both limitations. The proposed \crema\:and \cremb\:models, in fact, are fully probabilistic in nature and allow for the generation of network configurations characterized by continuous weights. Remarkably, for what concerns the binary estimation step, only the \emph{marginal probability distributions} $\{f_{ij}\}_{i,j=1}^N$ describing the behaviour of the random variables $\{a_{ij}\}_{i,j=1}^N$ are needed, a result that holds true irrespectively from the algorithm employed to derive the set of coefficients above.

Although it may be argued that the observations above hold true for the continuous version of the DECM as well, let us notice that its applicability is limited by the amount of information required to solve it, i.e. the knowledge of both the out- and in-degree sequences - a piece of information that is practically never accessible. On a more practical level, the numerical resolution of the \crema\:and \cremb\:models is much less costly than the numerical resolution of the DECM. Moreover, our framework allows us to further simplify the problem of finding a numerical solution of the system of eqs. (\ref{eq:sys_a1}), by providing a recipe to solve \emph{rescaled} versions of it: such a recipe can be employed to simplify calculations whenever a solution of the system above cannot be easily found at the considered scale (see also the Appendix).

The comparison between our two competing models, then, reveals that the best performance is achieved by the \cremb\:that is the clear winner both in terms of accuracy and simplicity of implementation - as it does not require the resolution of any system of equations; even more so, each parameter of the \cremb\:model can be computed \emph{independently} from the others, thus making the entire procedure parallelizable. To sum up, in order to achieve a fast and efficient reconstruction of weighted networks, we recommend the use of the \cremb\:model both in case of full uncertainty about the network topology and if the existence of some links is certain.

The codes to run both the \crema\:and the \cremb\:versions of our method are freely available at \cite{repository}.

\section*{Declarations}

\subsection*{Availability of Data and Materials}

Data concerning the World Trade Web are described in K. S. Gleditsch, Expanded trade and GDP data, \emph{Journal of Conflicts Resolution} {\bf 46}, 712-724 (2002) and can be found at the address http://privatewww.essex.ac.uk/~ksg/exptradegdp.html. Data concerning e-MID cannot be shared because of privacy issues preventing them from being publicly available. The codes implementing the conditional reconstruction algorithm can be found in \cite{repository}.

\subsection*{Competing Interests}

The authors declare no competing financial interests.

\subsection*{Authors Contributions}

F.P., T.S. and D.G. developed the method. F.P. performed the analysis. F.P., T.S. and D.G. wrote the manuscript. All authors reviewed and approved the manuscript.

\subsection*{Acknowledgements}

T.S. acknowledges support from the EU project SoBigData-PlusPlus (grant num. 871042). D.G. acknowledges support from the Dutch Econophysics Foundation (Stichting Econophysics, Leiden, the Netherlands) and the Netherlands Organization for Scientific Research (NWO/OCW).

\clearpage
\newpage

\appendix
\widetext

\section{The relationship between entropy and likelihood}

\paragraph*{Standard case.} Let us revise the relationship between Shannon entropy and likelihood in the standard case. The maximization of Shannon entropy

\begin{equation}\label{eq_app:S}
S(\mathcal{W})=-\sum_{\mathbf{A}\in\mathbb{A}}\int_{\mathbb{W}_\mathbf{A}}Q(\textbf{W})\log Q(\textbf{W})d\textbf{W}
\end{equation}
under the set of constraints $\vec{C}(\mathbf{W})$ leads to recover the functional \cite{gabrielli2018grand} form 

\begin{equation}
Q_{\la}(\mathbf{W})=\frac{e^{-H_{\la}(\mathbf{W})}}{Z_{\la}}
\end{equation}
where $H_{\la}(\mathbf{W})=\sum_{\alpha} \lambda_{\alpha}C_{\alpha}(\mathbf{W})$ is a linear combination of the constraints and $Z_{\la}=\int_{\mathbb{W}}e^{-H_{\la}(\mathbf{W})}d\mathbf{W}$ is the partition function. Upon substituting $Q_{\la}(\mathbf{W})$ into the Lagrangean function

\begin{equation}
\mathscr{L}(Q,\la)=S(\mathbf{\mathcal{W}})+\mu\left(1-\sum_{\mathbf{A}\in\mathbb{A}}\int_{\mathbb{W}_\mathbf{A}}Q(\textbf{W})\right)+\sum_{\alpha}\lambda_{\alpha}\left(C_{\alpha}^*-\langle C_{\alpha}\rangle\right)
\end{equation}
(with $\langle C_\alpha\rangle=\sum_{\mathbf{A}\in\mathbb{A}}\int_{\mathbb{W}_\mathbf{A}}Q(\textbf{W})C_{\alpha}(\mathbf{W})$) one recovers the expression

\begin{equation}
\mathscr{L}(Q(\la),\la)=\log Z_{\la}+\sum_{\alpha}\lambda_{\alpha}C_{\alpha}^*=-\ln Q_{\la}(\mathbf{W}^*)
\end{equation}
the last passage being valid for any graph $\mathbf{W}^*$ such that $\vec{C}(\textbf{W}^*)=\vec{C}^*$; in other words, the equation above states that the Lagrangean functional is ``minus'' the likelihood function of any graph $\mathbf{W}^*$ on which the constraints $\vec{C}(\textbf{W}^*)$ assume the values $\vec{C}^*$.

Notice that in case the Lagrangean functional is evaluated in $\la^*$, i.e. the parameters values ensuring that $\langle\vec{C}_{\la^*}\rangle=\vec{C}^*$, the result

\begin{equation}
\mathscr{L}(Q(\la^*),\la^*)=S(\la^*)=-\ln Q_{\la^*}(\mathbf{W}^*)
\end{equation}
is recovered.\\

\paragraph*{Conditional case.} In case the functional

\begin{equation}\label{eq_app:S2}
S(\mathcal{W}|\mathcal{A})=-\sum_{\mathbf{A}\in\mathbb{A}}P(\mathbf{A})\int_{\mathbb{W}_\mathbf{A}}Q(\textbf{W}|\mathbf{A})\log Q(\textbf{W}|\mathbf{A})d\textbf{W}
\end{equation}
is, instead, considered, the expression $Q_{\la}(\mathbf{W}|\mathbf{A})=\frac{e^{-H_{\la}(\mathbf{W})}}{Z_{\mathbf{A},\la}}$ is recovered, with obvious meaning of the symbols. Upon substituting $Q_{\la}(\mathbf{W}|\mathbf{A})$ into the Lagrangean function defined in eq. (\ref{eq:lagr}) one recovers the expression

\begin{equation}
\mathscr{L}(Q(\la),\la)=\sum_{\mathbf{A}\in\mathbb{A}}P(\mathbf{A})\log Z_{\mathbf{A},\la}+\sum_{\alpha}\lambda_{\alpha}C_{\alpha}^*=-\mathcal{G}(\la)
\end{equation}
the last passage being valid for any graph $\mathbf{W}^*$ such that $\langle\vec{C}\rangle^*=\vec{C}(\mathbf{W}^*)=\vec{C}^*$. Notice that in case the Lagrangean functional is evaluated in $\la^*$, i.e. the parameters values ensuring that $\langle\vec{C}_{\la^*}\rangle=\vec{C}^*$, the result

\begin{equation}
\mathscr{L}(Q(\la^*),\la^*)=S(\la^*)=-\sum_{\mathbf{A}\in\mathbb{A}}P(\mathbf{A})\ln Q_{\la^*}(\langle\mathbf{W}\rangle^*|\mathbf{A})=-\mathcal{G}(\la^*)
\end{equation}
is recovered.

\section{The binary reconstruction step: deriving $P(\mathbf{A})$}\label{app:P}

The Conditional Reconstruction Method works for any choice of $P(\mathbf{A})$. Here we derive two possible alternatives, to be selected according to the available information. The first possibility is deriving $P(\mathbf{A})$ by maximizing Shannon entropy 

\begin{equation}
S(\mathcal{A})=-\sum_{\mathbf{A}\in\mathbb{A}}P(\mathbf{A})\log P(\mathbf{A}),
\end{equation}
under the constraints represented by the out- and in-degree sequences $\{k_i^{out}\}_{i=1}^N$, $\{k_i^{in}\}_{i=1}^N$. This is the DBCM model \cite{squartini2018reconstruction} inducing a probability reading

\begin{equation}
P(\mathbf{A})=\prod_{j\neq i}p_{ij}^{a_{ij}}(1-p_{ij})^{1-a_{ij}}
\end{equation}
with $p_{ij}=\frac{x^{out}_ix^{in}_j}{1+x^{out}_ix^{in}_j}$. Unfortunately, this model is not viable for network reconstruction as the information about the degrees is practically never accessible. However, we can rest upon a certain approximately linear empirical linear relation generally found between the strengths and the Lagrange multipliers controlling for the degrees (i.e. $x^{out}_i\approx s^{out}_i\sqrt{a}$ and $x^{in}_i\approx s^{in}_i\sqrt{b}$ \cite{cimini2015systemic}) to define a more realistic model. This so-called `fitness ansatz' leads to our second model, whose linkage probability reads
\begin{equation}\label{eq:cimi}
p_{ij}=\frac{zs^{out}_is^{in}_j}{1+zs^{out}_is^{in}_j}
\end{equation}
(having defined $z\equiv\sqrt{ab}$), with the parameter $z$ tuned to reproduce the empirical link density: $\langle L\rangle=\sum_{j\neq i} p_{ij}=\sum_{j\neq i} \frac{zs^{out}_is^{in}_j}{1+zs^{out}_is^{in}_j}=L^*$. Equation (\ref{eq:cimi}) is the one characterizing the density-corrected Gravity Model \cite{cimini2015systemic}. Other possible choices for $P(\mathbf{A})$ are discussed in \cite{squartini2018reconstruction}.

\section{Solving the rescaled \crema\:problem}

Let us now prove that our framework easily allows one to find a solution of the system of eqs. (\ref{eq:sys_a1}) in case the sufficient statistics is rescaled, i.e. divided by an opportunely defined factor (e.g. $\kappa$). To this aim, let us consider the rescaled system

\begin{equation}\label{sysres}
\left\{\begin{array}{ll}
\sum_{j(\neq i)}\frac{f_{ij}}{\beta^{out}_i(\kappa)+\beta^{in}_j(\kappa)}&=\frac{s_i^{out^*}}{\kappa},\:\forall\:i\\
\sum_{j(\neq i)}\frac{f_{ji}}{\beta^{out}_j(\kappa)+\beta^{in}_i(\kappa)}&=\frac{s_i^{in^*}}{\kappa},\:\forall\:i
\end{array}\right.
\end{equation}
where the symbols $\vec{\beta}^{out}(\kappa)$ and $\vec{\beta}^{in}_i(\kappa)$ stress that the searched solutions are functions of the chosen rescaling parameter $\kappa$. A solution of the system above reads 

\begin{eqnarray}\label{sysres2}
\beta^{out^*}_i(\kappa)&=&\kappa\beta^{out^*}_i,\:\forall\:i\\
\beta^{in^*}_i(\kappa)&=&\kappa\beta^{in^*}_i,\:\forall\:i
\end{eqnarray}
as it can be proven upon substituting it back into eqs. (\ref{sysres}) and noticing that $\{\beta^{out^*}\}_{i=1}^N$ and $\{\beta^{in^*}\}_{i=1}^N$ are solutions of the system of eqs. (\ref{eq:sys_a1}). As our likelihood maximization problem admits a unique, global maximum, the prescription above allows us to easily identify it. Such a recipe turns out to enhance the chance of finding a solution to the system of eqs. (\ref{eq:sys_a1}) by solving a related problem at a more convenient scale.

\section{A ``golden standard'' for reconstruction models}

Let us notice that our framework allows us to define a sort of benchmark or ``golden standard'' for reconstruction models, defined by the assumptions that both the actual network topology and the entire set of weights are accessible. Upon considering that $\beta_{ij}=\frac{f_{ij}}{w_{ij}^*},\:\forall\:i\neq j$, implementing both assumptions leads to the conditional probability distribution

\begin{equation}
q_{ij}(w_{ij}^*|a_{ij}=1)=\frac{f_{ij}}{\langle w_{ij}\rangle}e^{-\frac{f_{ij}}{\langle w_{ij}\rangle}w_{ij}^*}=\frac{a_{ij}}{w_{ij}^*}e^{-\frac{a_{ij}}{w_{ij}^*}w_{ij}^*}=\frac{1}{e w_{ij}^*}
\end{equation}
further leading to 

\begin{equation}
\mathcal{G}(\la)=\ln Q(\mathbf{W}^*|\mathbf{A}^*)=\sum_{\{a_{ij}=1\}}(-1-\ln w_{ij}^*)=-L-\sum_{\{a_{ij}=1\}}\ln w_{ij}^*\nonumber\\
\end{equation}
i.e. to the maximum value of the likelihood attainable by a canonical model with local contraints, still preserving the strengths.

\section{Confidence intervals-based comparison of network models}

Both reconstruction models introduced in this manuscript induce pair-specific weight distributions: this allows a confidence interval to be defined around the expected value of each link weight. Since the procedure described below is valid for both the \crema\:and the \cremb\:models, let us consider a generic exponential distribution whose parameter is $\beta_{ij}$.

In order to analytically estimate the searched confidence interval $[w^-, w^+]$, let us solve the equation

\begin{equation}
\int_{w^-}^{\langle w_{ij}|a_{ij}=1\rangle}\beta_{ij}e^{-\beta_{ij}w_{ij}}dw_{ij}=q^-
\end{equation}
inverting which we find the left bound $w^-$; upon noticing that $\langle w_{ij}|a_{ij}=1\rangle=\frac{1}{\beta_{ij}}$ the result

\begin{equation}
w^-=-\frac{\ln[e^{-\beta_{ij}\langle w_{ij}|a_{ij}=1\rangle}+q^-]}{\beta_{ij}}=-\frac{\ln[e^{-1}+q^-]}{\beta_{ij}}
\end{equation}
is obtained. Analogously, the second equation to be solved is

\begin{equation}
\int_{w=\langle w_{ij}|a_{ij}=1\rangle}^{w^+}\beta_{ij}e^{-\beta_{ij}w_{ij}}dw_{ij}=q^+
\end{equation}
leading to the result
\begin{equation}
w^+=-\frac{\ln[e^{-\beta_{ij}\langle w_{ij}|a_{ij}=1\rangle}-q^+]}{\beta_{ij}}=-\frac{\ln[e^{-1}-q^+]}{\beta_{ij}}.
\end{equation}

Thus, upon fixing the desired confidence levels $q^-$ and $q^+$, the confidence interval $[w^-, w^+]$ accompanying the estimation of the conditional expected weight $\langle w_{ij}|a_{ij}=1\rangle$ is recovered. Generally speaking, such a confidence interval is not symmetric, given the peculiar form of the underlying probability distribution (i.e. the exponential one). 

As the comparison shown in fig. \ref{fig4} reveals, the \cremb\:model outperforms the \crema\:model in providing reliable estimates of actual weights.

\section{Deriving the continuous DECM}

The continuous DECM is obtained by maximizing the entropy $S(\mathcal{W})$ in eq.~\eqref{eq_app:S}
under the constraints represented by the out- and in-degree sequences and the out- and in-strength sequences:

\begin{equation}\label{eq:decm_sys}
\left\{\begin{array}{ll}
\langle k_i^{out}\rangle&=\sum_{\mathbf{A}\in\mathbb{A}}\int_{\mathbb{W}_\mathbf{A}}Q(\mathbf{W})k_i^{out}(\mathbf{W})d\mathbf{W}={k_i^{out}}^*\\
\langle k _i^{in}\rangle&=\sum_{\mathbf{A}\in\mathbb{A}}\int_{\mathbb{W}_\mathbf{A}}Q(\mathbf{W})k_i^{in}(\mathbf{W})d\mathbf{W}={k_i^{in}}^*\\
\langle s_i^{out}\rangle&=\sum_{\mathbf{A}\in\mathbb{A}}\int_{\mathbb{W}_\mathbf{A}}Q(\mathbf{W})s_i^{out}(\mathbf{W})d\mathbf{W}={s_i^{out}}^*\\
\langle s_i^{in}\rangle&=\sum_{\mathbf{A}\in\mathbb{A}}\int_{\mathbb{W}_\mathbf{A}}Q(\mathbf{W})s_i^{in}(\mathbf{W})d\mathbf{W}={s_i^{in}}^*
\end{array}\right.
\end{equation}

The continuous DECM is defined by a probability distribution reading $Q(\mathbf{W})=\frac{e^{-H(\mathbf{W})}}{Z}$ where

\begin{equation}
H(\mathbf{W})=\sum_{j\neq i}\left[(\alpha_i^{out}+\alpha_j^{in})\Theta(w_{ij})+(\beta_i^{out}+\beta_j^{in})w_{ij}\right]
\end{equation}
and

\begin{eqnarray}
Z&=&\sum_{\mathbf{A}\in\mathbb{A}}\int_{\mathbb{W}_\mathbf{A}}e^{-H(\mathbf{W})}d\mathbf{W}\nonumber\\
&=&\prod_{j\neq i}\int_{0}^{\infty}\left[\delta(w_{ij}-0)+\Theta(w_{ij})\right] e^{-(\alpha_i^{out}+\alpha_j^{in})\Theta[w_{ij}]-(\beta_i^{out}+\beta_j^{in})w_{ij}}dw_{ij}\nonumber\\
&=&\prod_{j\neq i}\left[1+e^{-(\alpha_i^{out}+\alpha_j^{in})}\int_{0}^{\infty}e^{-(\beta_i^{out}+\beta_j^{in})w_{ij}}dw_{ij}\right]\nonumber\\
&=&\prod_{j\neq i}\left[1+\frac{e^{-(\alpha_i^{out}+\alpha_j^{in})}}{\beta_i^{out}+\beta_j^{in}}\right]
\end{eqnarray}
($d\mathbf{W}$ stands for $\prod_{j\neq i}dw_{ij}$, in the first passage) which finally leads to

\begin{equation}\label{eq:q-cont-full_A}
Q(\mathbf{W})=\prod_{j\neq i}q_{ij}(w_{ij})=
\prod_{j\neq i}\frac{\left(x_i^{out}x_j^{in}\right)^{\Theta(w_{ij})}e^{-(\beta_i^{out}+\beta_j^{in})w_{ij}}}{1+x_i^{out}x_j^{in}/(\beta_i^{out}+\beta_j^{in})}.
\end{equation}

As we said in the main text, the \crema\:model admits the DECM as a particular case. In fact,

\begin{equation}
q_{ij}(w_{ij})=p_{ij}^{\text{DECM}}(\beta_i^{out}+\beta_j^{in})e^{-(\beta_i^{out}+\beta_j^{in})w_{ij}}
\end{equation}
for any positive weight and with $p_{ij}^{\text{DECM}}$ representing the probability that a link pointing from $i$ to $j$ exists. Indeed, the functional forms of the expected weights under the \crema\:and the DECM models coincide as well: in fact, $\langle w_{ij}\rangle_{\text{DECM}}=\frac{p_{ij}^{\text{DECM}}}{\beta_i^{out}+\beta_j^{in}}$, with $p_{ij}^{\text{DECM}}=p_{ij}^{\text{DECM}}(\alpha_i,\alpha_j,\beta_i,\beta_j)$, the latter expression explicitly showing the joint role played by binary and weighted constraints in determining the topological structure of the network at hand ($p_{ij}$ is, in fact, defined by the Lagrange multipliers associated with both the out- and in-degrees and the out- and in-strengths of nodes $i$ and $j$.
\end{document}